\documentclass[trackchanges,twocolumn]{aastex701}

\usepackage{threeparttable}
\usepackage{array, boldline, makecell, booktabs}

\begin{document}

\title{Probing the Atmospheres of Young Long-Period Sub-Neptune Progenitors with ELT/ANDES}

\author[gname=Spandan,sname=Dash]{Spandan Dash}
\affiliation{Université Paris Cité, Institut de physique du globe de Paris, CNRS, F-75005 Paris, France}
\email[show]{dash@ipgp.fr}

\author[gname=Dwaipayan,sname=Dubey]{Dwaipayan Dubey}
\affiliation{Universitäts-Sternwarte, Fakultät für Physik, Ludwig-Maximilians-Universität München, Scheinerstr. 1, D-81679 München, Germany}
\affiliation{Exzellenzcluster `Origins’, Boltzmannstr. 2, D-85748 Garching, Germany}
\email{ddubey@usm.lmu.de; 2014dwaipayan@gmail.com}

\author[gname=Liton,sname=Majumdar]{Liton Majumdar}
\affiliation{Exoplanets and Planetary Formation Group, School of Earth and Planetary Sciences, National Institute of Science Education and Research, Jatni - 752050, Odisha, India}
\affiliation{Homi Bhabha National Institute, Training School Complex, Anushaktinagar, Mumbai 400094, India}
\email{liton@niser.ac.in; dr.liton.majumdar@gmail.com}

\begin{abstract}
High-resolution cross-correlation spectroscopy (HRCCS) has become a powerful ground-based technique for detecting and characterizing exoplanet atmospheres. While highly successful for ultra-hot and hot Jupiters, next-generation facilities such as ELT/ANDES will observe smaller and longer-period planets, including young sub-Neptunes and their progenitors. We investigate whether HRCCS with ELT/ANDES can robustly recover orbital parameters and atmospheric signals for the long-period sub-Neptunes V1298\,Tau\,b and TOI-451\,c. In long-period systems, the slow Doppler drift during a single night limits separation between planetary and telluric signals, increasing the risk of signal loss during detrending. We therefore quantify the impact of including out-of-transit exposures on signal recovery and parameter estimation. We simulate YJH-band transmission observations using the \texttt{Ratri} pipeline and analyze them with the HRCCS detrending and cross-correlation framework \texttt{Upamana}. For V1298\,Tau\,b, injected atmospheric models are consistent with HST, Spitzer, and JWST constraints. For TOI-451\,c, we explore sub-solar to super-solar C/O ratios to test compositional sensitivity. Incorporating out-of-transit exposures significantly improves detectability, provided detrending effects are consistently propagated to the template spectra prior to cross-correlation. Without this step, orbital parameters can deviate from injected values and detection significance decreases. For V1298\,Tau\,b, $>4\sigma$ detections of H$_2$O, H$_2$S, and CO are achievable at $\lesssim$10 hours (minimum 2 nights, cloud-free scenario). For TOI-451\,c, distinguishing sub-solar and solar from super-solar C/O requires  $\sim$17 hours (minimum 4 nights). HRCCS with ELT/ANDES will therefore be a key tool for atmospheric characterization of young, long-period sub-Neptunes in the ELT era.

\end{abstract}



\section{Introduction}
Starting from the first detection of orbital motion in transmission of the hot-Jupiter HD 209458\,b in 2010 \citep{snellen2010orbital}, the high-resolution\footnote{R = $\lambda/\Delta\lambda > 25000$} cross-correlation spectroscopy (HRCCS) technique has become a complementary tool to lower-resolution space-based observations to characterise exoplanetary atmospheres. This technique takes advantage of the fact that the exoplanetary signal during a night of observation is Doppler-shifted appreciably in wavelength space over time (a few km\,s$^{-1}$) to stand out from the photon-dominated noise (in which it is embedded) left over after all other contributions to the total flux received at the spectrograph, which either do not shift in wavelength space or shift at velocities lower than the instrumental velocity resolution, have been removed. Although initially focussed on detecting single or multiple molecular and atomic species in the atmospheres of hot and ultra-hot Jupiters \citep{brogi2012signature,rodler2012weighing,de2013detection,birkby2013detection,brogi2014carbon,brogi2016rotation,hawker2018evidence,cabot2019robustness}, the development of likelihood frameworks in \citet{brogi2019retrieving}, \citet{gandhi2019hydra} and \citet{gibson2020detection} has led to enabling of atmospheric retrievals by coupling with Bayesian parameter estimators, and thus pushed this technique into the realm of characterisation of exoplanetary atmospheres. This means that both lower-resolution space-based and high-resolution ground-based datasets for target exoplanet systems can now be analysed simultaneously to provide more precise constraints on atmospheric parameters \citep{brogi2019retrieving,gandhi2019hydra,kasper2022unifying,boucher2023co,smith2024combined}. More recently, the technique has also been utilized to either successfully detect molecules in sub-Neptune atmospheres \citep{basilicata2024gaps}, or to showcase the potential to detect molecules in the atmospheres of such exoplanets from injection tests while ultimately not detecting an equivalent signal from observed data \citep{lafarga2023hot,dash2024constraints,grasser2024peering,parker2025limits,pelaez-torres_tighter_2025}. Among super-Earths, 55\,Cnc\,e has been a consistent target for high-resolution studies \citep{ehrenreich2012hint,zhang2021no,ridden2016search,esteves2017search,jindal2020characterization,tabernero2020horus,deibert2021near,keles2022pepsi,rasmussen2023nondetection}, but they have not resulted in any positive detections so far, while observability studies using coupled interior-atmosphere models for their spectral templates showcase that mineral atmospheres should be detectable in emission using current ground-based spectrographs like CARMENES \citep{dash2025detectability}. This trend of non-detection has also held true for other rocky exoplanets like GJ 486\,b \citep{ridden2023high} and GJ 1132\,b \citep{palle2025exploring}.
\\
\\
One of the most important steps of HRCCS analysis, apart from the accuracy of the template model being cross-correlated against, is the detrending procedure where all contributions to the total flux other than the Doppler-shifted exoplanet signal are removed. The two most common approaches used in literature today to perform this procedure are singular value decomposition (SVD) \citep[first used for this purpose in ][]{de2013detection} and SYSREM \citep[first used for this purpose in ][]{birkby2013detection}. Both approaches identify the highest contributions to the flux received at spectrographs and then remove those contributions by either subtracting out the contributions to fluxes due to the first few singular/eigenvectors alone or by normalising the total flux received by those contributions. However, this process is not an unbiased approach and leads to a phase-dependant impact (termed from here on as artefacts) on line depths (attenuation) and line wing shapes (distortion) on an injected signal in the high-resolution dataset being detrended \citep{brogi2019retrieving,gibson2020detection,pino2022gaps,meech2022applications,gibson2022relative,cheverall2024feasibility,dash2024constraints}. Additionally, it also leads to production of these artefacts around the injected signal, extending to all the exposures with an exoplanet signal in emission and transmission datasets \citep{brogi2019retrieving, dash2024constraints}, as well as in the out-of-transit exposures for transmission datasets brought on by the detrending algorithm over-correcting at these phases due to the presence of an injected signal in the in-transit phases \citep[][ also please see Appendix \ref{reprocesseffect}]{cheverall2024feasibility,dash2024constraints,palle2025ground}.
\\
\\
Presumably the detrending algorithm also has a similar effect on the actual exoplanet signal embedded in the dataset, which is why the same detrending procedure is also performed on the template exoplanet signal that is used for the cross-correlation step \citep[for a fast method for SYSREM, please see][]{gibson2022relative}. The presence of out-of-transit artifacts has been shown to enhance detection significance when all detrended exposures (both in and out-of-transit) are used in the cross-correlation step, compared to using in-transit exposures alone \citep{dash2024constraints,palle2025ground}. These results raise the tantalizing possibility that the HRCCS may remain effective even for long-period planets in the ELT era, provided that some out-of-transit exposures are obtained during the night of observation \citep{cheverall2024feasibility,palle2025ground}. To evaluate this scenario, we apply an SVD-based detrending algorithm to realistic ELT/ANDES simulations of the long-period sub-Neptunes V1298\,Tau\,b and TOI-451\,c. V1298\,Tau\,b has an orbital period of 24.139 $\pm$ 0.001 days \citep{david2019four} and TOI-451\,c has an orbital period of 9.192 $\pm$ 0.000064 days \citep{kokori2023exoclock}. Considering both planets allows us to test the performance of the SVD-based detrending scenario across a wide range of orbital periods.
\\
\\
Young sub-Neptunes such as V1298\,Tau\,b and TOI-451\,c represent an emerging class of planets whose atmospheres still retain signatures of their early formation and mass-loss histories. Both planets are part of the JWST Cycle 3 program GO 5959 (KRONOS: Keys to Revealing the Origin and Nature Of sub-neptune Systems), which targets young systems to probe how low-mass atmospheres form, cool, and undergo mass loss. Their moderate equilibrium temperatures and inflated radii make them ideal laboratories for assessing whether next-generation high-resolution spectrographs on ELT can probe atmospheric chemistry beyond the hot-Jupiter regime. V1298\,Tau\,b, orbiting a young ($<$ 30 Myr) pre-main-sequence K-type T-Tauri star, offers a rare benchmark with existing space-based spectra \citep{oh2017vizier,luhman2018stellar,david2019warm,barat2024first,barat2025metal}. Initially classified as a warm Jupiter \citep{david2019warm}, it was later proposed to be an inflated sub-Neptune progenitor \citep{kubyshkina2020coupling,owen2020constraining,vach2024occurrence} that has not yet contracted or lost a significant fraction of its envelope. HST/WFC3 observations revealed a clear, extended atmosphere with a strong 1.4  $\mu$m H$_2$O feature and a large scale height \citep{barat2024first}. Follow-up observations with JWST/NIRSpec G395H revealed evidence for a haze-free, sub-solar C/O atmosphere with absorption features from H$_2$O, CO$_2$, CO, CH$_4$, SO$_2$, and OCS \citep{barat2025metal}. The combined HST and JWST retrievals indicate a low metallicity, close to solar values, which is unusual when compared to mature sub-Neptunes that generally exhibit much higher heavy-element enrichments \citep{thorngren2016mass}. In contrast, TOI-451\,c orbits a young solar-type star with an estimated age of $\sim$120 Myr \citep{newton2021tess}, placing it at a later but still formative stage of sub-Neptune evolution. Its shorter orbital period and higher equilibrium temperature relative to V1298\,Tau\,b make it an important complementary target for exploring comparative exoplanetology through atmospheric chemistry and observability analysis. Despite the absence of current space-based atmospheric constraints, its young age, inflated radius, and proximity to the host star suggest that TOI-451\,c may likewise represent a sub-Neptune progenitor and hence, an exciting test case for evaluating the sensitivity of HRCCS to young, low-mass planets under more extreme irradiation conditions.
\\
\begin{table*}[t]
\centering
\caption{Adopted stellar and planetary parameters for the V1298\,Tau\,b and TOI-451\,c systems used in the atmospheric structure and disequilibrium chemistry calculations. Stellar properties, orbital parameters, and planetary masses and radii are taken from the literature as indicated.}
\begin{threeparttable}
\begin{tabular*}{\textwidth}{@{\extracolsep{\fill}}lcccc@{}}
\noalign{\smallskip}
\toprule
& \multicolumn{2}{@{}c@{}}{V1298\,Tau\,b} & \multicolumn{2}{@{}c@{}}{TOI-451\,c} \\\cmidrule{2-3}\cmidrule{4-5}%
Parameters & Values & References & Values & References\\
\noalign{\smallskip}
\hlineB{3.5}
\noalign{\smallskip}
$M\mathrm{_{star}}$ [$M_{\odot}$] & 1.26$^{0.06}_{-0.06}$ & \cite{finociety2023monitoring} & 0.95$^{0.020}_{-0.020}$ & \cite{newton2021tess} \\
\noalign{\smallskip}
$R\mathrm{_{star}}$ [$R_{\odot}$] & 1.43$^{0.03}_{-0.03}$ & \cite{finociety2023monitoring} & 0.88$^{0.0032}_{-0.032}$ &   \cite{newton2021tess}\\
\noalign{\smallskip}
$T\mathrm{_{star}}$ [K] & 4941$^{+31}_{-16}$ & \cite{finociety2023monitoring} & 5550$^{56}_{-56}$ & \cite{newton2021tess}  \\
\noalign{\smallskip}
$P_\mathrm{rot, \star}$ [days] & 2.91$^{0.005}_{-0.005}$ & \cite{finociety2023monitoring}  & 5.1$^{0.1}_{-0.1}$ & \cite{newton2021tess}  \\
\noalign{\smallskip}
$v_\mathrm{sys}$ [kms$^{-1}$] & 16.15$^{0.38}_{-0.38}$ & \cite{david2019warm} & 19.87$^{0.12}_{-0.12}$ & \cite{newton2021tess}  \\
\noalign{\smallskip}
$M\mathrm{_{planet}}$ [$M\mathrm{_{Jup}}$] & 0.0412$^{0.0167}_{-0.0167}$ & \cite{livingston2026young}  & 0.012$^{0.010}_{-0.008}$ & \cite{barragan2026mass}  \\
\noalign{\smallskip}
$R\mathrm{_{planet}}$ [$R\mathrm{_{Jup}}$] & 0.840$^{0.051}_{-0.051}$ & \cite{livingston2026young} & 0.277$^{0.012}_{-0.012}$ &   \cite{newton2021tess}\\
\noalign{\smallskip}
T$_{eq, \mathrm{Planet}}$ [K] & 677$^{22}_{-22}$  & \cite{david2019four} & 875$^{+13}_{-11}$ & \cite{newton2021tess}  \\
\noalign{\smallskip}
log($g$)\tnote{a} [cgs] & 3.28 & -- & 3.00 & --  \\
\noalign{\smallskip}
$a$ [au] & 0.169$^{0.0026}_{-0.0026}$ & \cite{david2019four} & 0.082$^{+0.0033}_{-0.0036}$ & \cite{newton2021tess}  \\
\noalign{\smallskip}
Period [days] & 24.14$^{0.0018}_{-0.0018}$& \cite{david2019four} & 9.19$^{+0.00006}_{-0.00010}$ & \cite{newton2021tess}  \\
\noalign{\smallskip}
Distance [pc] & 108.2$^{0.7}_{-0.7}$ & \cite{stassun2019revised} & 123.74$^{+0.39}_{-0.38}$ & \cite{stassun2019revised}  \\
\noalign{\smallskip}
Transit Duration [hours] & 6.42$^{0.13}_{-0.13}$ & \cite{david2019four} & 3.56$^{0.038}_{-0.038}$ & \cite{newton2021tess}  \\
\noalign{\smallskip}
\noalign{\smallskip}
\hline
\hline
\end{tabular*}
\begin{tablenotes}
\item[a] The values are taken from the \href{https://exo.mast.stsci.edu/}{Exo.MAST} database, as these parameters have not been reported in recent peer-reviewed studies.
\end{tablenotes}
\end{threeparttable}
\label{tab:systems}
\end{table*}
\\
Building on this observational context, we generate forward models for both V1298\,Tau\,b and TOI-451\,c to compare their detectability with HRCCS using ELT/ANDES. For V1298\,Tau\,b, the availability of HST and JWST/NIRSpec G395H spectra provides a direct reference for validating our high-resolution templates. In contrast, TOI-451\,c currently lacks comparable atmospheric observations. We therefore explore a controlled set of atmospheric scenarios for this planet by assuming solar metallicity and three chemically plausible C/O ratios (solar, sub-solar, and super-solar). This design captures the expected diversity of atmospheric compositions for young sub-Neptunes and enables us to quantify the sensitivity of HRCCS to chemical variation. Together, this two-planet framework provides a systematic test-bed for assessing whether SVD-based HRCCS pipelines can robustly recover atmospheric signals across a range of orbital and chemical conditions relevant to ELT-ANDES. We outline our modeling and analysis framework and discuss how each component supports our objective of evaluating SVD-based HRCCS performance for young sub-Neptunes. Section \ref{sec:methods} describes our methodology, including the generation of chemically self-consistent atmospheric structures, the computation of line-by-line high-resolution transmission spectra, and the synthesis of ANDES observations used for the HRCCS analysis. Section \ref{sec:results} presents the detectability tests for both planets, compares and discusses the results obtained with the presence and absence of a reprocessing step before cross-correlation analysis (for V1298\,Tau\,b), showcases how atmospheric chemistry can influence the recovery of injected signals and whether it is possible to differentiate between different atmospheric chemistry scenarios using our synthetic nights (for TOI-451\,c). In Section \ref{sec:conclusion}, we conclude and state the implications of our findings on the feasibility of detecting and characterizing the population of young, low-mass, and long-period exoplanets with ANDES.

\begin{figure*}
\centering
  \includegraphics[width=0.9\textwidth]{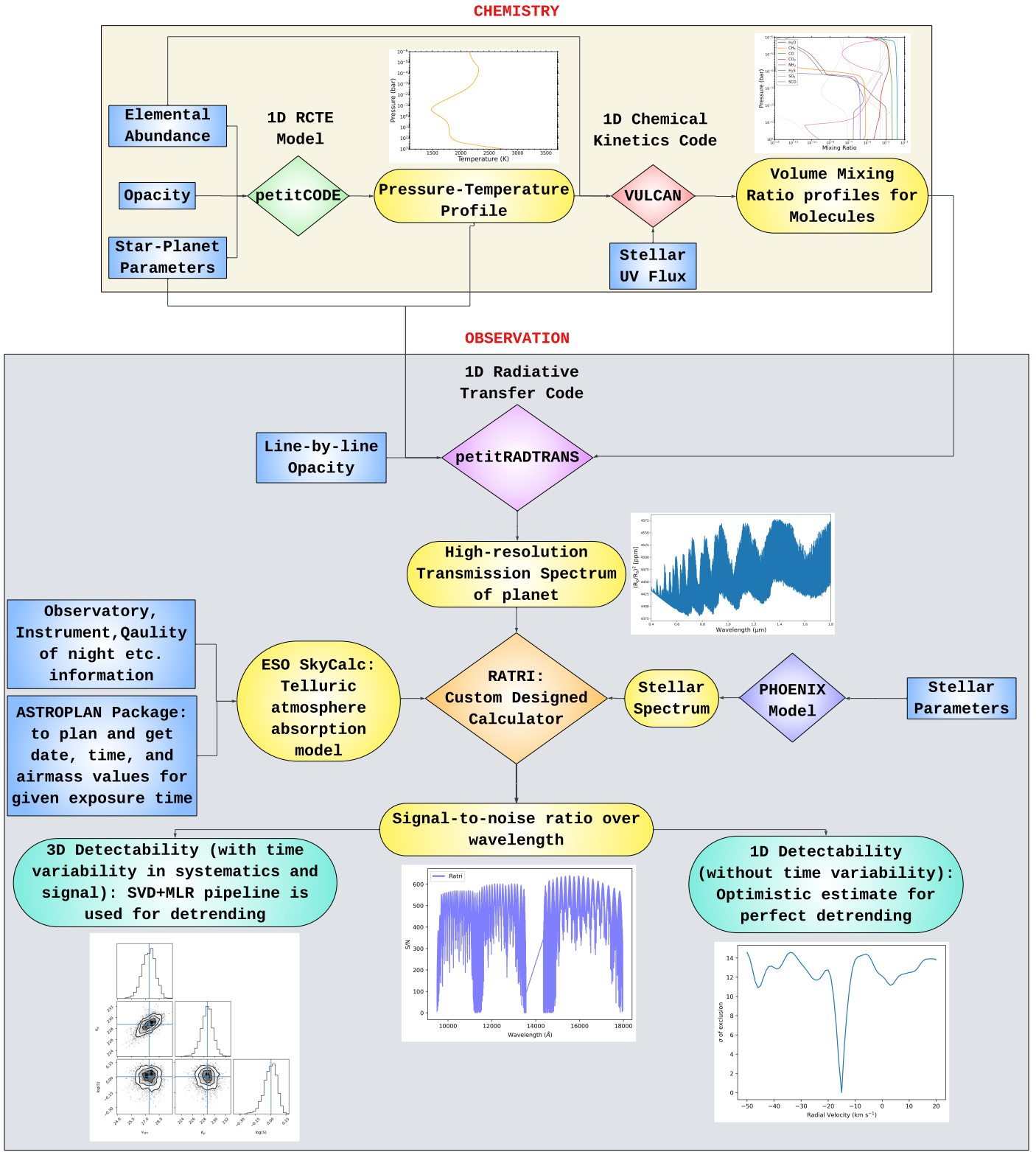}
  \caption{Overview of the modeling and analysis workflow used in this study. The Chemistry module (top) combines elemental abundances, opacities, and star–planet parameters to compute 1D pressure–temperature profiles using 1D radiative-convective-thermochemical equilibrium model petitCODE, which are then coupled to the 1D chemical kinetics code, VULCAN, to produce volume mixing ratio profiles under stellar UV irradiation. The Observation module (middle) uses petitRADTRANS to generate line-by-line high-resolution transmission spectra, which are propagated through the \texttt{Ratri} ELT/ANDES simulator together with \texttt{PHOENIX} stellar models, telluric transmission from \texttt{ESO SkyCalc}, and observatory and airmass information obtained via \texttt{astroplan}. The resulting wavelength-dependent signal-to-noise ratios feed into two detectability pathways: (1) a 1D ``optimistic” case assuming perfect detrending, and (2) a full 3D detectability test incorporating time-variable systematics and applying the SVD + MLR detrending pipeline. This end-to-end framework is used to assess HRCCS detectability for young sub-Neptunes in this study.}
  \label{fig:flowchart}
\end{figure*}

\section{Methods}
\label{sec:methods}

Figure~\ref{fig:flowchart} provides the overview of the modeling pipeline followed in this work. Our framework consists of three main stages. First, we compute one-dimensional radiative–convective–thermochemical equilibrium structures using \texttt{petitCODE} \citep{molliere2015model,molliere2017modeling}, and then derive disequilibrium abundance profiles with the 1D chemical kinetics model \texttt{VULCAN}\footnote{\href{https://github.com/exoclime/VULCAN}{https://github.com/exoclime/VULCAN}}  \citep{tsai2017vulcan,tsai2021comparative}. Second, these profiles are used to generate high-resolution line-by-line transmission spectra with \texttt{petitRADTRANS}\footnote{\href{https://petitradtrans.readthedocs.io/en/latest/}{https://petitradtrans.readthedocs.io/en/latest/}} (version 3.2.0) \citep{molliere2019petitradtrans,molliere2020retrieving,2024JOSS....9.5875N}. Third, the spectra are propagated through the high-resolution observation simulator \texttt{Ratri} \citep{dash2025detectability}, which incorporates stellar \texttt{PHOENIX} stellar models \citep{husser2013new}, telluric transmission, and observatory conditions. Ultimately, the resulting time series observations are analysed with the SVD + MLR (multi-linear regression)-based HRCCS pipeline \texttt{Upamana} \citep{dash2024constraints,dash2025detectability}. The subsections below describe each component of this workflow.


\subsection{Integrated Modeling of Atmospheric Structure and Disequilibrium Chemistry}
\label{sec:chemistry}

\begin{figure*}
    \centering  
         
        \begin{minipage}[b]{0.31\textwidth}
                \includegraphics[width=\textwidth]{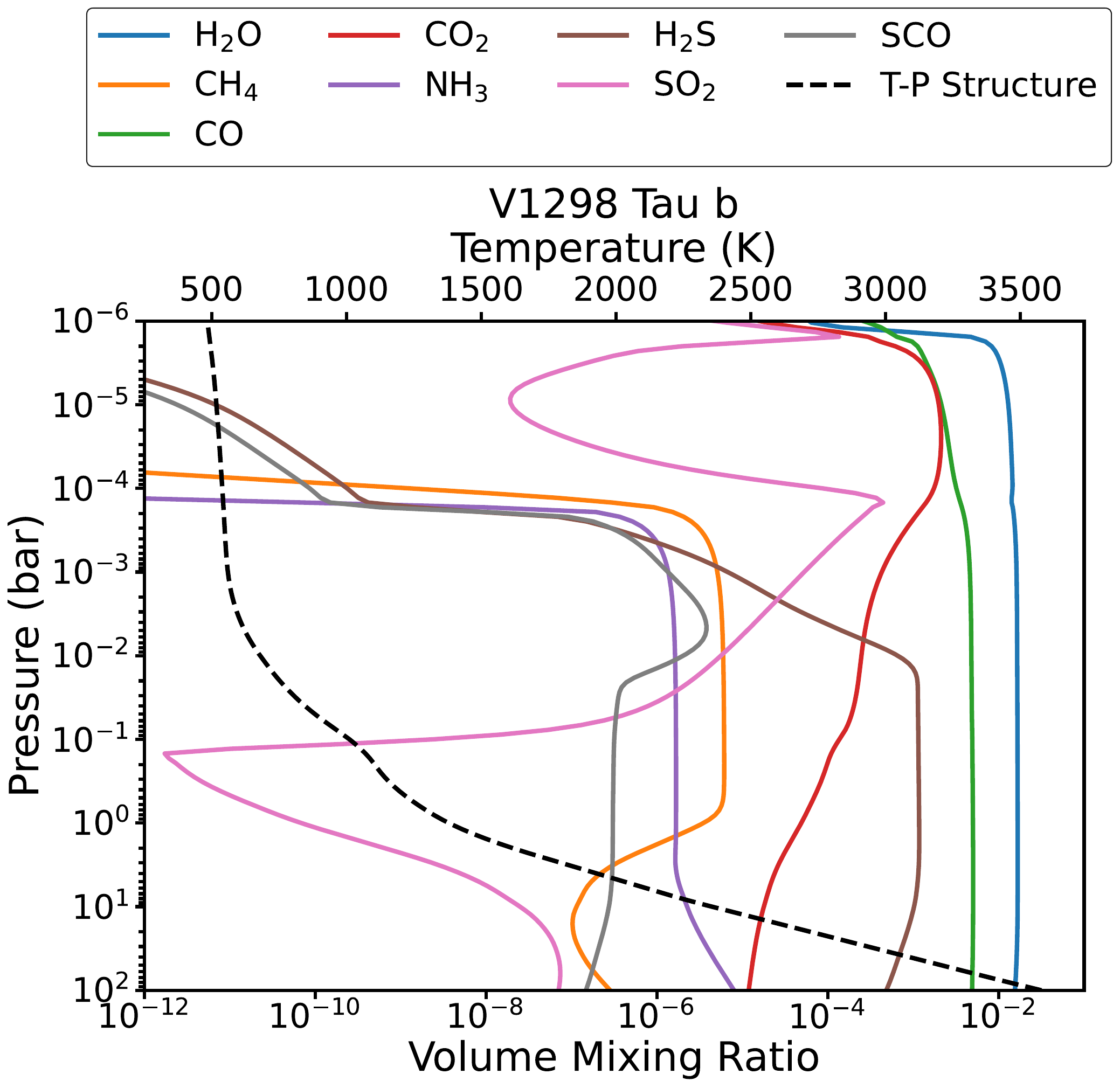}
        \end{minipage}\\
         \vspace{0.5cm}
        \begin{minipage}[b]{0.32\textwidth}
            \includegraphics[width=\textwidth]{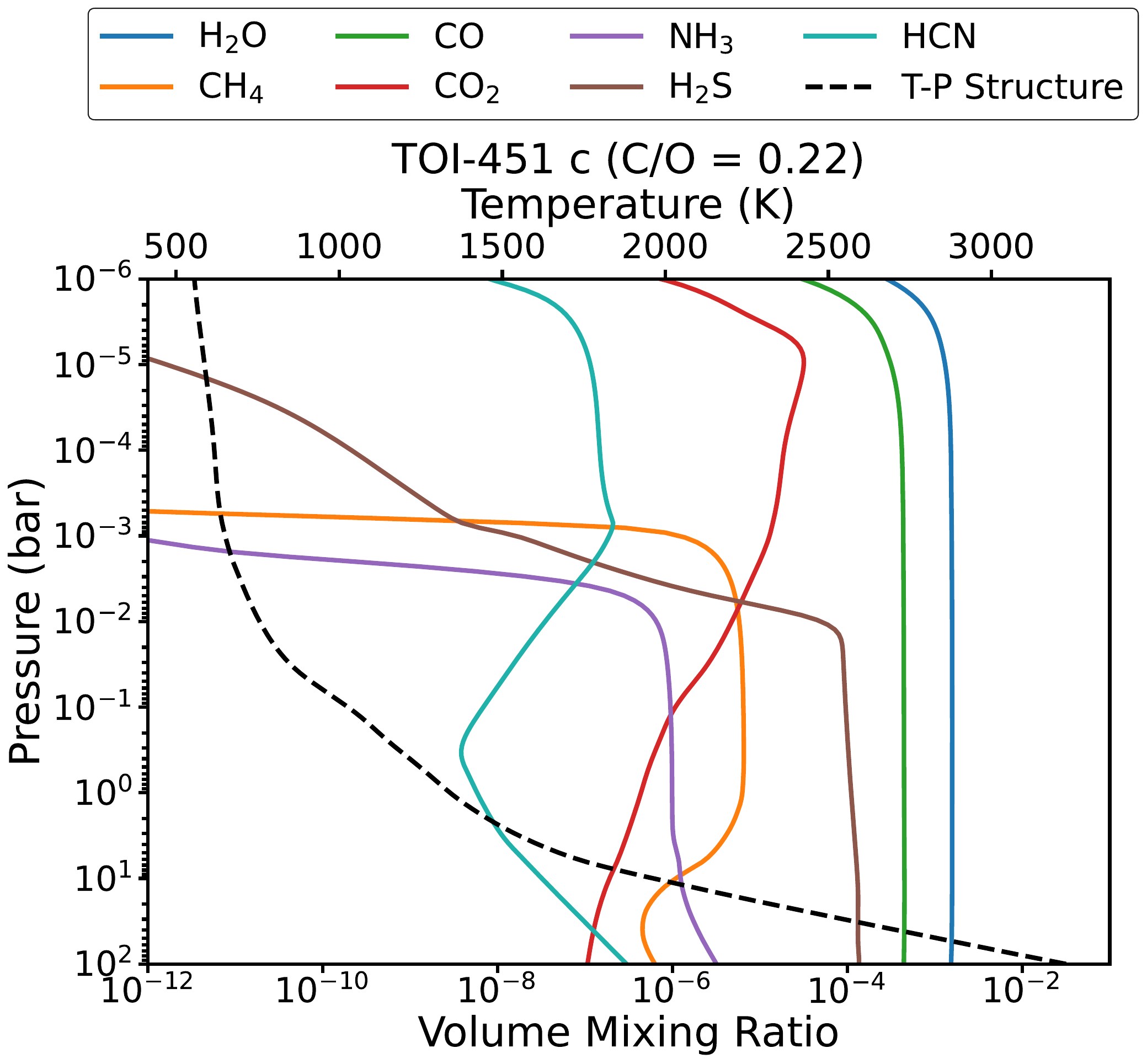}
        \end{minipage}
        \begin{minipage}[b]{0.32\textwidth}
                \includegraphics[width=\textwidth]{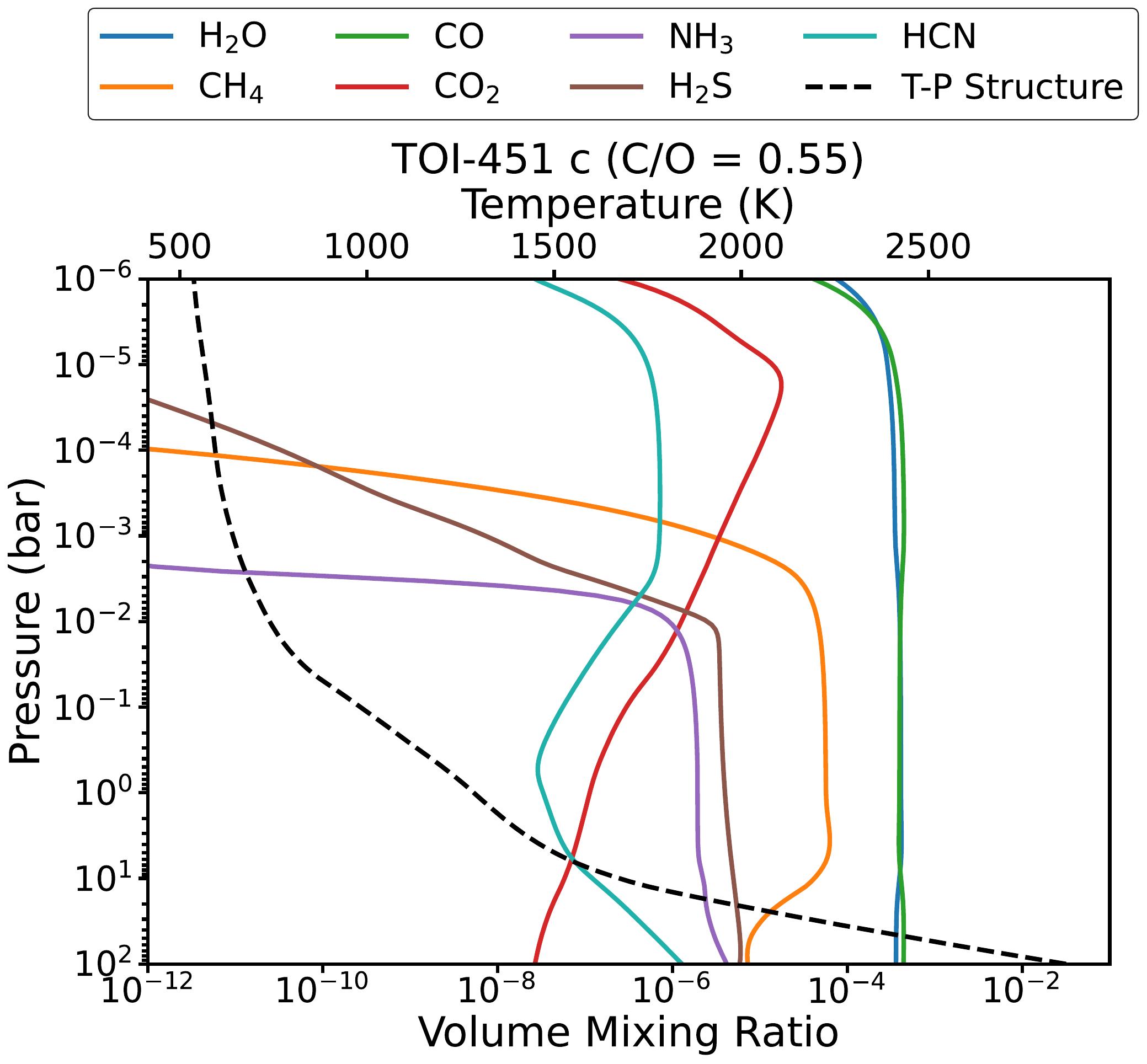}
        \end{minipage}
        \begin{minipage}[b]{0.32\textwidth}
        \includegraphics[width=\textwidth]{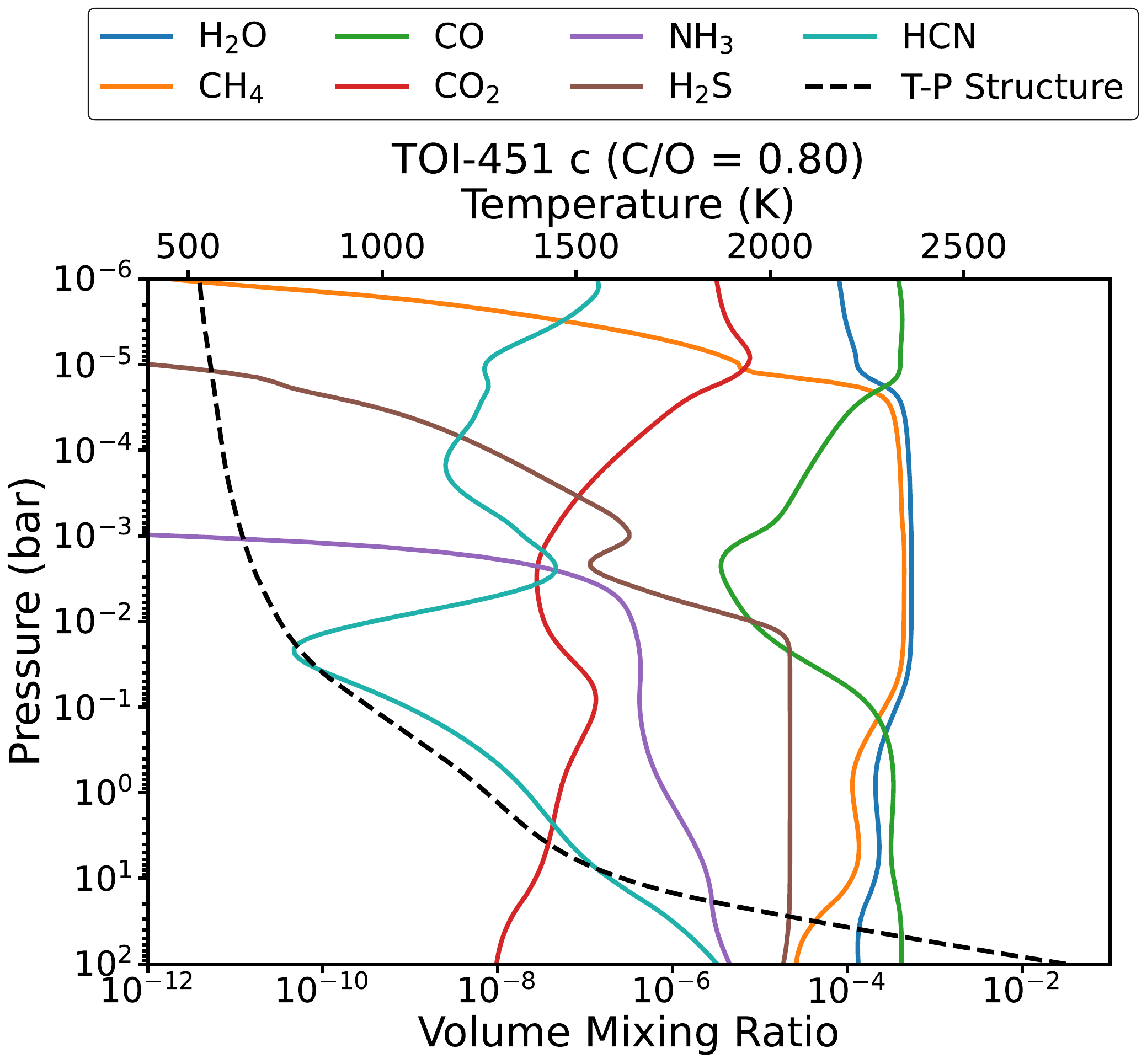}
        \end{minipage}
\caption{Vertical chemical abundance profiles for the dominant molecular species in the atmospheres of V1298\,Tau\,b (top panel) and TOI-451 c (bottom panels). The top panel shows the disequilibrium chemistry results for V1298\,Tau\,b computed using a self-consistent radiative–convective pressure–temperature profile and the star-planet parameters listed in Table \ref{tab:systems} The three bottom panels correspond to TOI-451 c models assuming sub-solar (C/O = 0.22), solar (C/O = 0.55), and super-solar (C/O = 0.80) elemental carbon-to-oxygen ratios at solar metallicity. All models include vertical mixing and photochemistry computed with \texttt{VULCAN}. The profiles illustrate the sensitivity of key species to elemental composition and disequilibrium processes across the observable atmosphere. Molecules shown are selected based on having average volume mixing ratios exceeding 10$^{-8}$ between the 1 bar and 0.1 mbar pressure levels, corresponding to the observable photospheric region.}
\label{fig:chemisrty}
\end{figure*}

The atmospheric chemistry for V1298\,Tau\,b and TOI-451\,c is carried out using a combined radiative–convective and disequilibrium-kinetics framework. We discretize each atmosphere among 160 layers over a pressure grid spanning $10^{2}$ to $10^{-7}$ bar. For each planet, we first compute a self-consistent pressure-temperature (PT) profile with \texttt{petitCODE}, which solves for radiative-convective and thermochemical equilibrium. and then use this structure as input to \texttt{VULCAN} to calculate disequilibrium abundance profiles.
\\
\\
The atmosphere is constructed with the following opacity sources relevant for warm sub-Neptune atmospheres: H$_2$O, CO, CO$_2$, CH$_4$, NH$_3$, HCN, H$_2$S, C$_2$H$_2$ (HITRAN; see \cite{gordon2022hitran2020}), PH$_3$ \citep{sousa2015exomol}, Na, K(VALD3; see \cite{piskunov1995vald}), TiO (ExoMol; \cite{mckemmish2019exomol}), and VO (ExoMol; \cite{bowesman2024exomol}). The treatment of Continuum opacity includes the collision-induced absorption (CIA) for H$_2$–H$_2$ and H$_2$–He pairs \cite{borysow1988collison,borysow1989collision,borysow1989collision2,borysow2002collision,richard2012new} and H$^{-}$ bound–free absorption \citep{2008oasp.book.....G}. This similar opacity treatment is also followed in recent studies \citep{dubey2023polycyclic,dubey2024comparative,dubey2025quantified}. We assume cloud-free atmospheres for both planets. For V1298\,Tau\,b, this choice is supported by the HST/WFC3 observation \citep{barat2024first}, which revealed a clear atmosphere with no detectable aerosol opacity. For TOI-451\,c, no observations exist to date, and adopting a cloud-free configuration keeps consistency in the comparison of chemical and high-resolution spectroscopic detectability across the two systems.
\\
\\
For V1298\,Tau\,b, we model the self-consistent PT profile by implementing the retrieved atmospheric parameters from \citet{barat2025metal}: carbon-to-oxygen ratio [C/O] = 0.23 and metallciity [M/H] = 10$^{1.05}$, and internal temperature [T$_{int}$] = 500 K. Elemental abundances are assigned using the metallicity and C/O ratio prescriptions of \cite{madhusudhan2012c,molliere2015model,woitke2018equilibrium,molaverdikhani2019cold1,dubey2023polycyclic,dubey2024comparative,dubey2025quantified}. Metallicity scales all heavy elements except H and He, and the C/O ratio is imposed by adjusting the oxygen abundance while carbon is held fixed. For TOI-451 c, since the atmosphere is not constrained from any observation, we compute different atmospheric structures assuming solar metallicity, a similar internal temperature of 500 K but spanning sub-solar, solar, and super-solar C/O ratios (C/O = 0.22, 0.55, and 0.80, respectively). This design captures the expected transition from H$_2$O-dominated to CO-dominated to CH$_4$-rich thermochemical regimes. 
\\
\begin{figure}
\centering
  \includegraphics[width=\columnwidth]{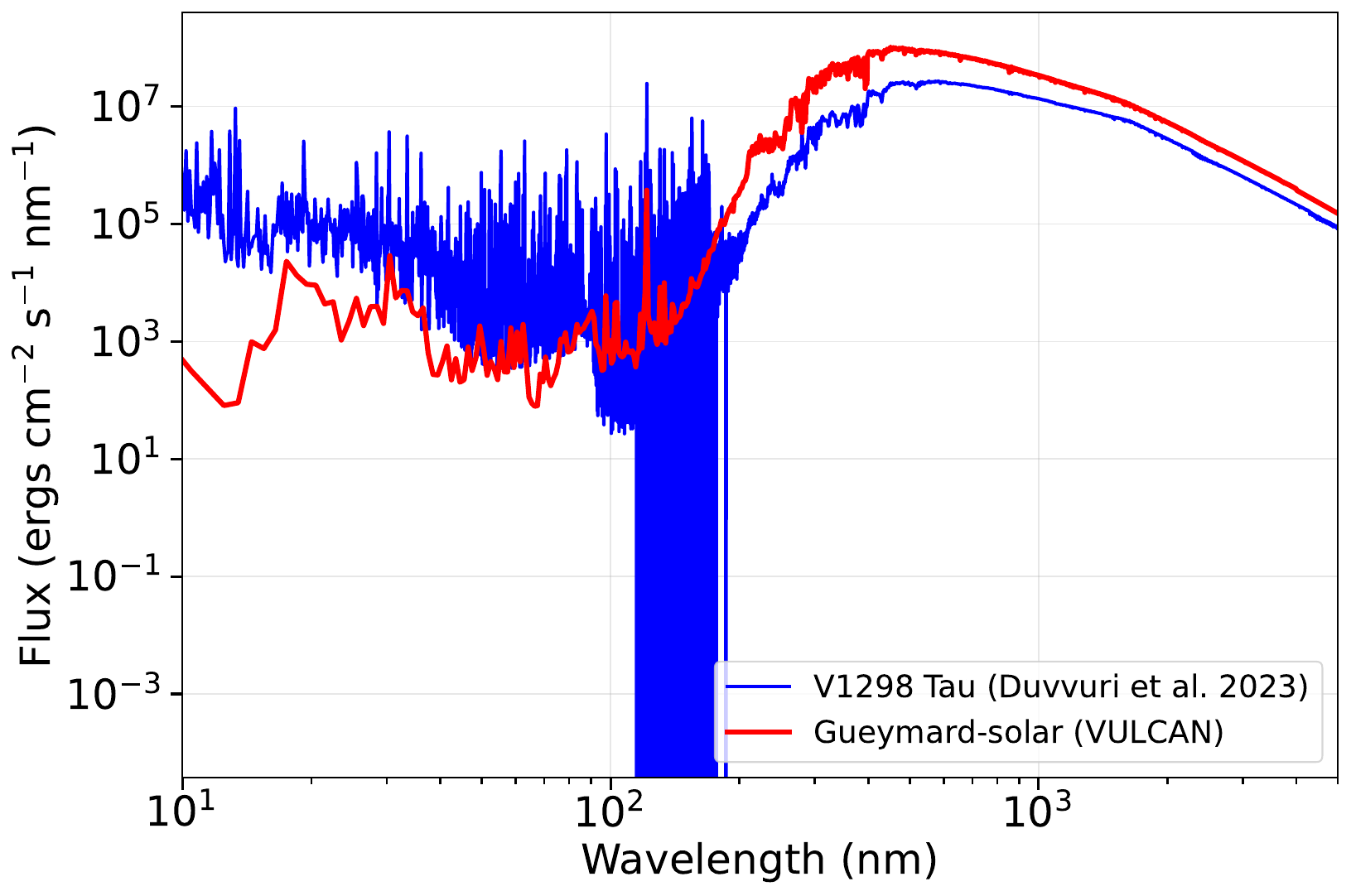}
  \caption{Comparison of the ultraviolet stellar flux spectra used to model photochemistry in the atmospheres of V1298\,Tau\,b and TOI-451\,c. The V1298\,Tau spectrum from \cite{duvvuri2023high} is rescaled to the stellar surface, whereas the ``\texttt{Gueymard-solar}'' spectrum provided with \texttt{VULCAN} is adopted as a proxy for TOI-451, whose effective temperature is comparable to that of the Sun.}
  \label{fig:uv_flux_comparison}
\end{figure}
\\
Once the PT profiles are computed self-consistently, we integrate them to \texttt{VULCAN} to simulate the atmospheric chemistry under the influence of molecular diffusion, atmospheric mixing (by the implementation of Eddy diffusion coefficient) and photochemistry. A few star-planet parameters, such as Star radius, planet radius, star-planet distance, and planet gravity, are essential for calculation of disequilibrium chemistry (see Table \ref{tab:systems}). For both cases, we solve the chemistry for the ``\texttt{S-N-C-H-O photo network}" listed in the VULCAN repository. For both planets, Eddy diffusion coefficient ($K_{zz})$ of 10$^{7}$~cm$^{2}$\,s$^{-1}$ is implemented following \cite{barat2025metal}. For V1298\,Tau\,b, the stellar ultraviolet (UV) spectrum used to model photochemistry in the upper atmosphere is taken from \cite{duvvuri2023high}. The spectrum reported in their study corresponds to the stellar flux as observed at Earth. To apply this spectrum in the context of the V1298\,Tau\,b atmosphere, we rescaled the observed flux to the surface of V1298\,Tau star using the measured distance to the system and the star radius. Since TOI-451 star has an effective temperature comparable to the Sun, we implement the ``\texttt{Gueymard-solar}" UV spectrum from the VULCAN repository. A comparison of the ultraviolet flux profiles adopted for V1298\,Tau and TOI-451 stars is shown in Figure \ref{fig:uv_flux_comparison}. The He/H ratio is held constant at 0.085, while all other elemental abundances are scaled consistently with the  metallicity and C/O ratio for each planet. Once everything is set, VULCAN computes the initial molecular abundances from its coupled equilibrium chemistry solver, FastChem\footnote{\href{https://github.com/exoclime/FastChem}{https://github.com/exoclime/FastChem}} \citep{stock2018fastchem}, and then the model is iterated until the molecules converge to a steady state.
\\
\\
The chemical abundance profiles of the dominant species for V1298\,Tau\,b and for three atmospheric scenarios of TOI-451\,c are shown in Figure \ref{fig:chemisrty}. For V1298\,Tau\,b, the resulting chemical profiles are broadly consistent with the one-dimensional radiative-convective-photochemical equilibrium (RCPE) grid-chemistry retrievals for the JWST NIRSpec G395H observations \citep{barat2025metal}. Notably, however, our chemistry model predicts sufficiently high abundances of $\mathrm{SO_2}$ in the photospheric region to reproduce the weak absorption feature near 4.05 $\mu$m. This $\mathrm{SO_2}$ signature was detected at a $4\sigma$ confidence level in free-chemistry retrievals but is not reproduced by the RCPE grid-chemistry retrieval in \cite{barat2025metal}. In contrast, the atmospheric composition of TOI-451\,c exhibits moderate variations as a function of the assumed C/O ratio. In particular, the vertical abundance profiles of $\mathrm{H_2O}$, $\mathrm{CH_4}$, CO, and $\mathrm{H_2S}$ show the strongest sensitivity to changes in elemental composition.
\\
\\
Although one of the TOI-451\,c cases adopts the same sub-solar C/O ratio as inferred for V1298\,Tau\,b, the resulting atmospheric chemistry is not identical. In particular, the sulfur chemistry differs, with sulfur-bearing species such as SO$_2$ remaining much less prominent in the TOI-451\,c models. This shows that the bulk elemental ratio alone is not sufficient to determine the observable chemical composition of these young sub-Neptune atmospheres. The differences arise from the combined effect of the adopted stellar irradiation environment and the bulk atmospheric properties used in the models, including the different metallicity assumptions and star-planet parameters. We therefore treat TOI-451\,c not as a direct chemical analogue of V1298\,Tau\,b, but as a distinct comparison case, and explore three C/O scenarios to test how chemically plausible diversity in young sub-Neptune atmospheres propagates into the high-resolution transmission spectra and HRCCS detectability analysis.

\subsection{Transmission spectra simulation from planet atmospheres}
\label{sec:spectra}

\begin{figure*}
    \centering  
         
        \begin{minipage}[b]{\columnwidth}
                \includegraphics[width=\columnwidth]{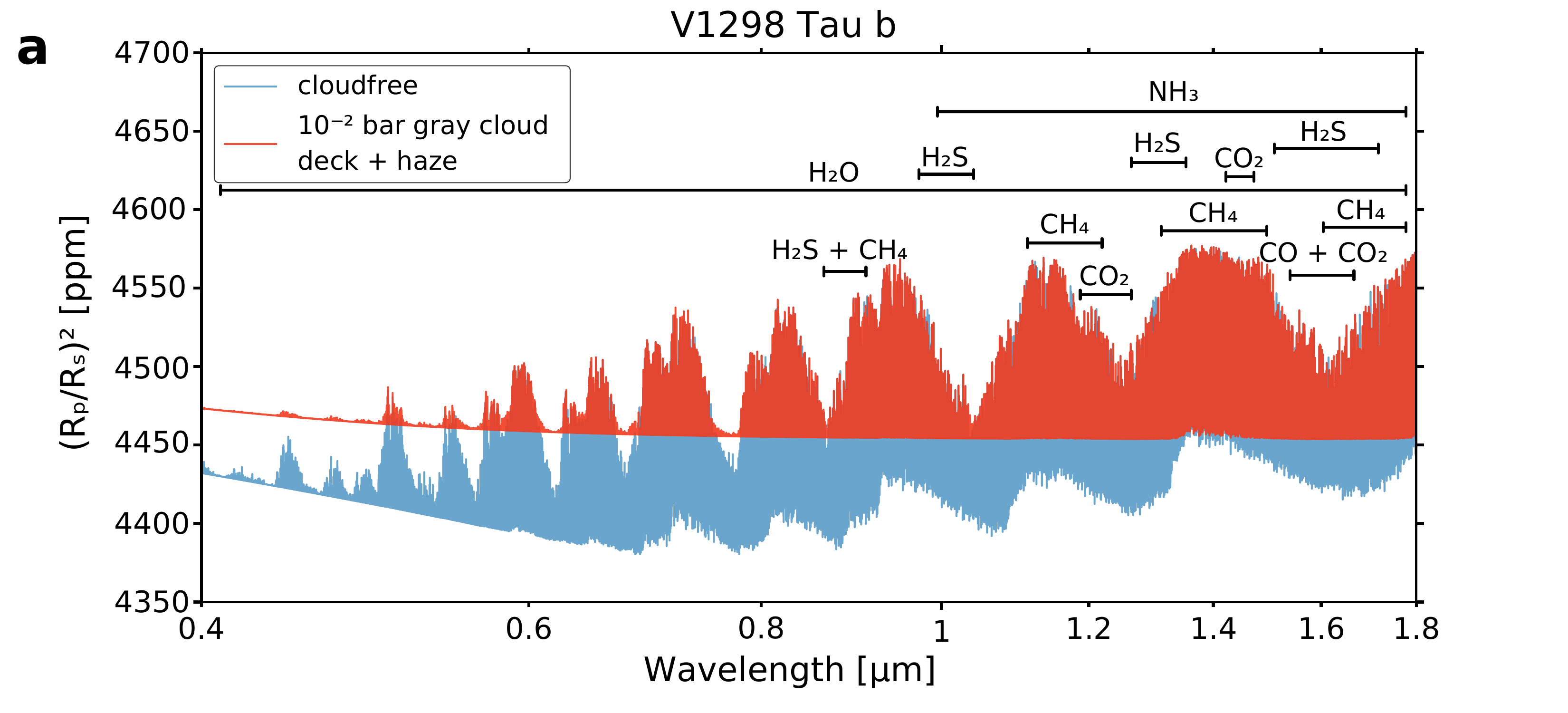}
        \end{minipage}\\
        \begin{minipage}[b]{\columnwidth}
            \includegraphics[width=\columnwidth]{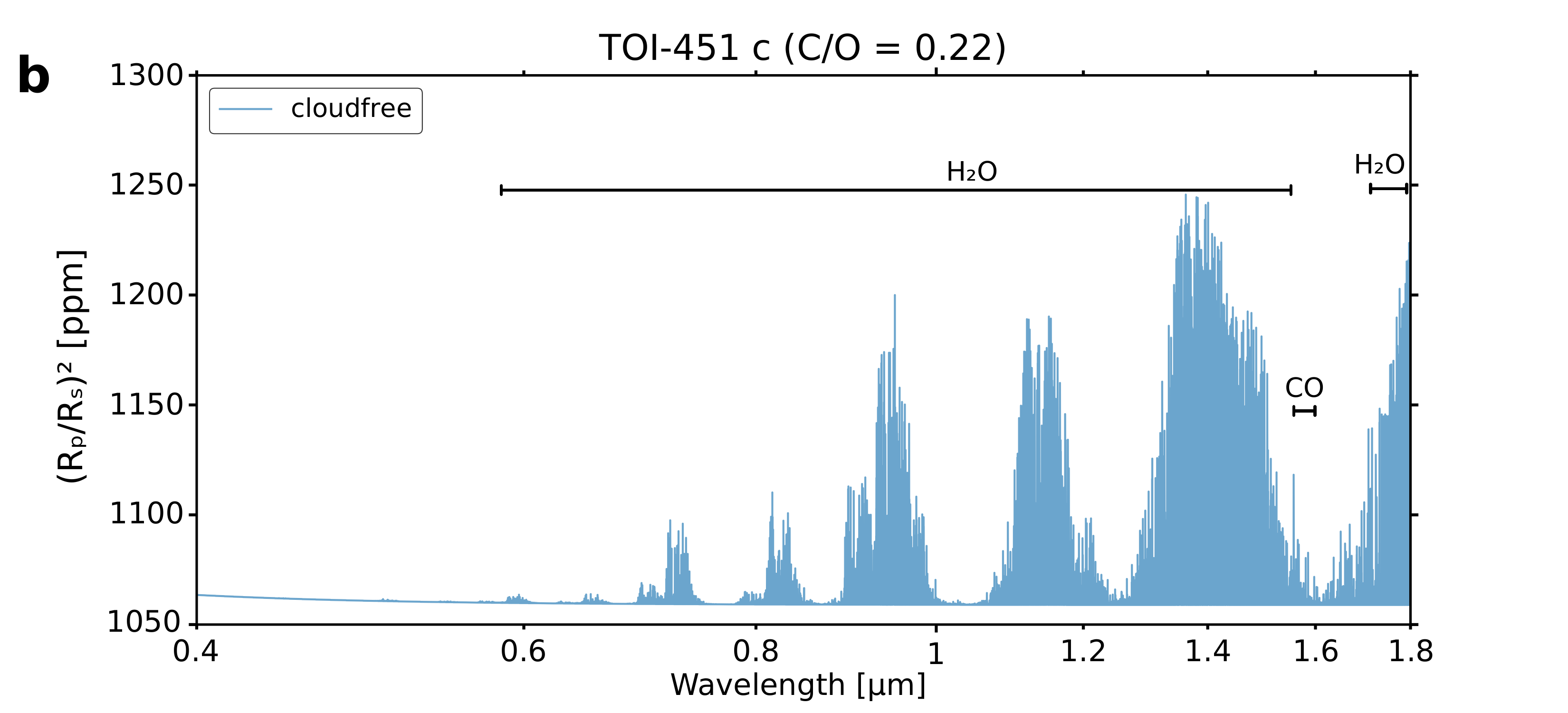}
        \end{minipage}
        \begin{minipage}[b]{\columnwidth}
                \includegraphics[width=\columnwidth]{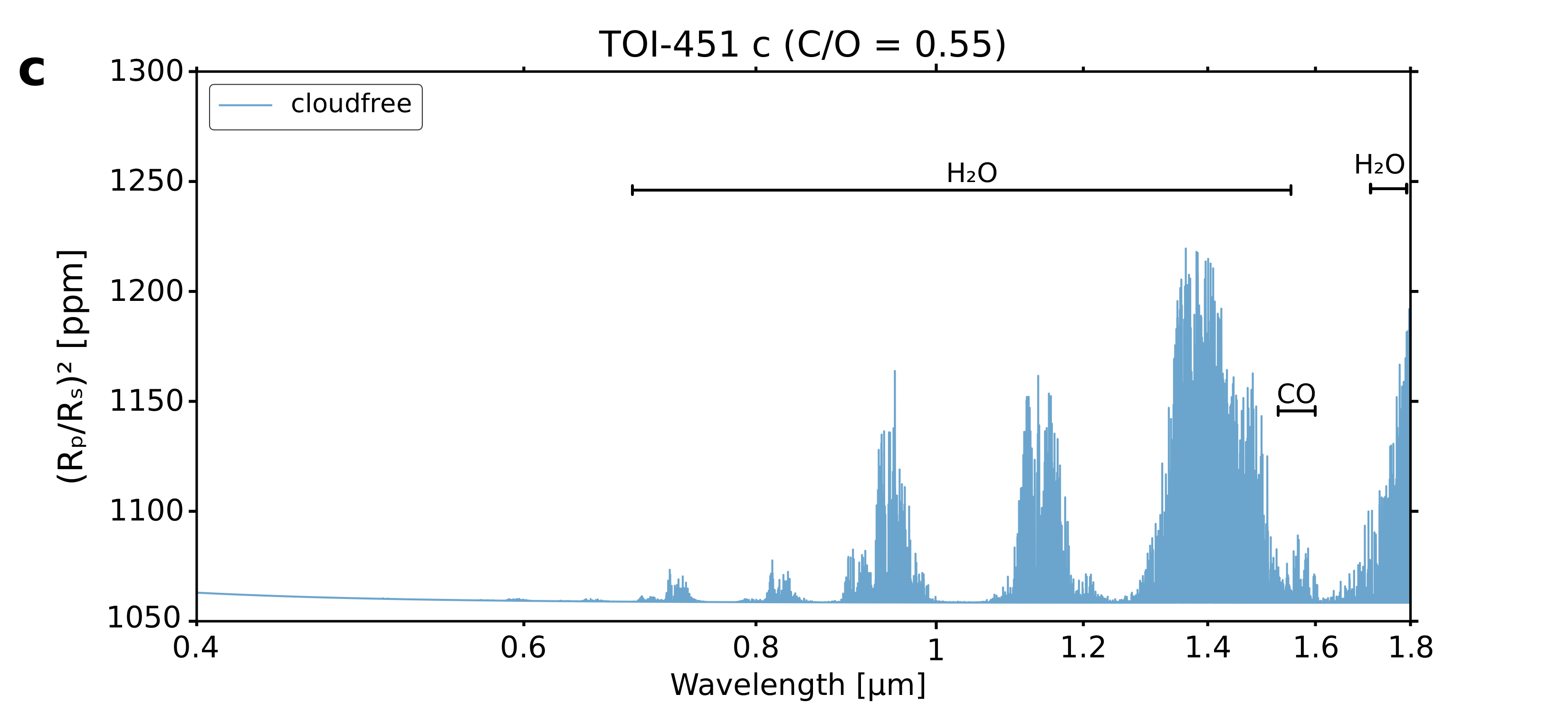}
        \end{minipage}
        \begin{minipage}[b]{\columnwidth}
        \includegraphics[width=\columnwidth]{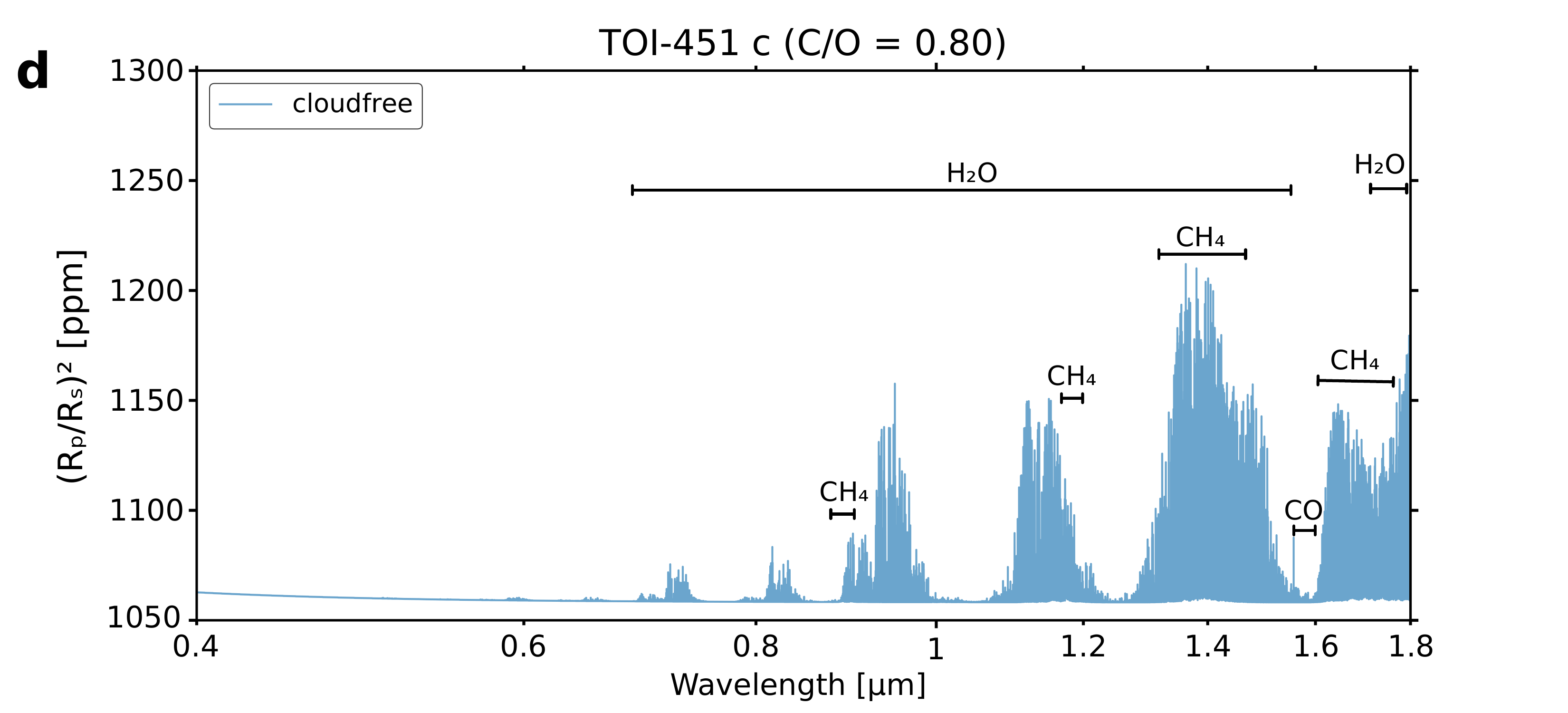}
        \end{minipage}
\caption{High-resolution transmission spectra simulated for V1298\,Tau\,b and TOI-451\,c at the resolving power of ELT/ANDES: (a) spectrum of V1298\,Tau\,b computed using disequilibrium chemistry and a self-consistent pressure–temperature profile, while panels; (b)-(d) correspond to TOI-451\,c assuming sub-solar (C/O = 0.22), solar (C/O = 0.55), and super-solar (C/O = 0.80) elemental carbon-to-oxygen ratios, respectively. For V1298\,Tau\,b, cloud-free and cloudy atmospheric configurations are investigated, where the latter includes a gray cloud deck at 0.01 bar and an enhanced Rayleigh-scattering haze. All spectra include collision-induced absorption and Rayleigh scattering scaled by a factor of 10 and are convolved with the instrumental line-spread function of ANDES. The contributions of individual molecular species to the total transmission spectrum are shown as a function of wavelength.}
\label{fig:spectra}
\end{figure*}

We compute transmission spectra for both planets by coupling time-dependent atmospheric chemistry with high-resolution, line-by-line radiative transfer. Molecular abundance profiles, obtained from VULCAN, are provided as inputs to the radiative-transfer framework petitRADTRANS\footnote{\href{https://petitradtrans.readthedocs.io/en/latest/}{https://petitradtrans.readthedocs.io/en/latest/}}
, where atmospheric transmission spectra are calculated relative to a reference pressure of $P_0 = 0.01$ bar. To investigate the influence of clouds on observable spectral features, we consider two idealized atmospheric configurations: a cloud-free atmosphere and a cloudy atmosphere that includes a gray cloud deck at 0.01 bar (indicated from JWST observations in \citealt{barat2025metal}) together with a haze-like component. In the gray cloud prescription, the atmosphere is assumed to be completely opaque below the cloud deck by assigning very large opacities to deeper layers. The haze component is implemented by scaling the Rayleigh scattering cross-section of the gas by a factor of 10.
\\
\\
The radiative-transfer calculations include molecular opacity from species that are chemically relevant in each atmosphere. For V1298\,Tau\,b, we consider $\mathrm{H_2O}$, $\mathrm{CH_4}$, CO, $\mathrm{CO_2}$, $\mathrm{NH_3}$, $\mathrm{H_2S}$, $\mathrm{SO_2}$, and SCO, while for TOI-451\,c we include $\mathrm{H_2O}$, $\mathrm{CH_4}$, CO, $\mathrm{CO_2}$, $\mathrm{NH_3}$, HCN, and $\mathrm{H_2S}$. Opacity data are taken from the high-resolution petitRADTRANS database or computed following the approach described in \cite{dubey2025quantified}. Molecules are retained only if their mean volume mixing ratio within the photospheric region (pressures between 1 bar and 0.1 mbar) exceeds $\mathrm{10^{-8}}$; species with lower abundances are excluded as they do not contribute appreciably to the transmission spectra. In addition to molecular absorption, we include collision-induced absorption from $\mathrm{H_2}$-$\mathrm{H_2}$ and $\mathrm{H_2}$-He interactions, as well as Rayleigh scattering by $\mathrm{H_2}$ and He. 
\\
\\
The intrinsic resolving power of the molecular opacity data is $R = 10^{6}$. In the radiative-transfer calculations, the ELT spectral resolution is incorporated through a combination of opacity sampling and instrumental convolution. The native opacities are first rebinned to the resolving power of the ELT ($R_{\rm ELT} = 10^{5}$) using an opacity-downsampling factor of 10, as implemented in petitRADTRANS. The resulting spectra are then convolved with a Gaussian line-spread function, characterised by a standard deviation $\mathrm{\sigma_{LSF} = R/(2\sqrt{ln2}})$, before being mapped onto the wavelength grids of ANDES. This step is essential for capturing realistic line broadening and preventing artificial enhancement of spectral contrast. To guarantee proper sampling of the instrumental resolution, we enforce the Nyquist criterion on all wavelength grids, such that $\lambda/\Delta\lambda$ $>$ 2$\lambda/\Delta\lambda_{\rm LSF}$. Here, $\lambda$ denotes the wavelength at which the opacities in petitRADTRANS are defined, $\Delta\lambda$ is the wavelength spacing between adjacent grid points, and $\Delta\lambda_{\rm LSF}$ corresponds to the full width at half maximum of the spectrograph’s line-spread function. 
\\
\\
The resulting high-resolution transmission spectra for V1298\,Tau\,b and TOI-451\,c are shown in Figure \ref{fig:spectra}. For V1298\,Tau\,b, spectra is presented for both cloud-free and cloudy atmospheric configurations. Individual molecular contributions are shown to indicate the dominant opacity sources as a function of wavelength. For TOI-451\,c, only cloud-free atmospheres are considered, and the comparison across different assumed C/O ratios illustrates how variations in elemental composition modify the relative strengths of H$_2$O, CO, CH$_4$, and H$_2$S-dominated spectral regions.

\subsection{Extending \texttt{Ratri} to ELT-ANDES}
\label{sec:ratri}
\texttt{Ratri} \footnote{\href{https://github.com/dashspandan/Ratri}{https://github.com/dashspandan/Ratri}} is an open-source python-based simulator for synthesizing nights of observations at any given date and time using high-resolution ground-based spectrographs \citep{dash2025detectability}. It was initially constructed to synthesize observations using the four best performing spectrographs among the currently operating ground-based high-resolution spectrographs - CARMENES, GIANO, SPIRou and CRIRES+. In this study, we have extended this list to also include the ANDES spectrograph, which is scheduled to be mounted on the ELT in the future. We have now also explicitly included random sampling of a lognormal distribution of Precipitable Water Vapour (PWV) values during any simulated observation, with the mean value of the PWVs depending on the location of a spectrograph (or user-specified depending on the quality of night required). The framework to determine the amount of flux received at the ground while taking into account the time-varying telluric absorption through Earth's atmosphere remains the same as described in detail in \citet{dash2025detectability}. However, we have now used a telescope of diameter 39.2\,m to account for the larger collecting area of ELT. We have also included sky emission as an additional source of flux so that the resultant noise is not strictly source photon-noise, even though we operate \texttt{Ratri} in the photon-noise dominated regime in this study (peak SNR/resolution element per order $>$ 300 in the YJH band). Although it is already possible to do so with the current setup of the pipeline, we have not incorporated a treatment for read-out noise (ROD) and dark-current noise, because the CCD configuration for ANDES has not yet been finalized. Additionally, with the values of dark-current noise and ROD provided in \citet{dubey2024comparative} for ANDES, and assuming that 2 pixels are used per resolution element with an exposure time of 60\,s (as will be used for V1298\,Tau\,b in this study), the dark-current noise (i.e. the square root of the dark-current value) per resolution element is 0.36 $\mathrm{e^-}$s, and the ROD noise is 4.2 $\mathrm{e^-}$s. Both of these terms are negligible compared to the source photon-noise value which is $>$ 300 $\mathrm{e^-}$s at its lowest and about 600 $\mathrm{e^-}$s at its highest, except in the most telluric saturated pixels which will all be masked out during the detrending procedure, and hence can be safely neglected for our feasibility simulations. We use information about efficiencies and telescope throughput available online from the ANDES Exposure Time Calculator (ANDES-ETC) \citep{sanna2024andes,palle2025ground}\footnote{ETC available at: https://andes.inaf.it/instrument/exposure-time-calculator/}. However, not all values were provided at the ANDES operating resolution of 100,000, so we interpolated all values on to a standardized wavelength grid of 100,000 for each wavelength band in which ANDES is expected be operational (UBV, RIZ, YJH and K) using nearest neighbour interpolation. We compare the time-invariant (for a single airmass value of 1.2) Signal-to-Noise Ratio (SNR) per resolution element output of \texttt{Ratri} against the output from the ANDES exposure time calculator for 60\,s of exposure of V1298\,Tau in Appendix \ref{andesratricomp} and find reasonable agreement across all the available wavelength bands, with \texttt{Ratri} providing slightly higher values overall across all bands  with the highest values of \texttt{Ratri} $\sim$ 1.1 $\times$ the highest values from ANDES-ETC, while reproducing the same trend of SNR variation overall. Any differences could be because \texttt{Ratri} works with a distance based framework, uses \texttt{PHOENIX} stellar models as default, and uses modelled telluric absorption from ESO SkyCalc \citep{noll2012atmospheric,jones2013advanced}, whereas the ANDES-ETC works with a stellar magnitude based framework, uses Pickles stellar spectra library \citep{pickles1998stellar}, and uses the TAPAS framework \citep{bertaux2014tapas} to calculate atmospheric transmission.
\\
\\
The ANDES-ETC does not provide spectral orders versus pixel array grids to mimic the CCD configuration, while information about the starting and ending wavelengths for each order is provided. Thus, for ease of calculation, we decided to interpolate all efficiencies and throughput values on to the YJH band CCD configuration of CARMENES, which covers almost the same wavelength range as ANDES, and is already included in \texttt{Ratri}. We only needed to scale up the size of the resolution element to be the same as that of ANDES to be able to find very similar SNR/resolution element values that we find for a standardized grid through the process outlined in the preceding paragraph. We note that CARMENES has 28 spectral orders and 4096 pixels, while the order-wise information in ETC-ANDES shows that ANDES can have double the number of orders. Since the detrending procedure itself is not spectrograph dependent, especially for ultra-stable spectrographs like CARMENES and (hopefully) for ANDES as well, we do not expect much differences in the processing step between our approach here and with future realistic observations. However, the wavelength coverage within orders and the number of pixels/resolution element can differ across the instruments, with CARMENES being sparser in comparison, which means that the trend of SNR/resolution element can vary. Consequently, the finer wavelength coverage in ANDES can result in higher values of the cross-correlation function (with more line cores of the template model available to be included in this step) compared to what we show in this study. Thus, the results of this study should be taken as a conservative estimate compared to results from actual observations in the future. Having spectral orders versus pixel array grids makes it easy to synthesize flux cuboids of dimensions $n_\mathrm{orders}\times n_\mathrm{spectra}\times n_\mathrm{pixels}$, where $n_\mathrm{orders}$ is the number of spectral orders (28 for CARMENES), $n_\mathrm{spectra}$ is the number of exposures taken during the night of observation, and $n_\mathrm{pixels}$ is the number of pixels (4096 for CARMENES) as a proxy for the wavelength. For finding the number of exposures, based on total observation time, we also assume a read-out time of 34\,s between exposures (same as CARMENES), which is slightly higher than the 30\,s used in \citet{palle2025ground}.
\\
\\
Once the flux cuboids have been prepared, the time-varying exoplanet signal is injected into the cuboids. For this, we first normalize the template spectrum prepared in Section \ref{sec:spectra} to their continuum, and then this normalized spectrum is Doppler-shifted in time corresponding to the exoplanetary parameters found in literature (see Table \ref{tab:systems}). The Doppler-shifted spectrum is injected into each of the exposures through the equation:
\begin{equation}
    F_{\lambda,\star+\mathrm{P}} = F_{\lambda,\star}\bigg{[}1-\bigg{(}\frac{R_{\lambda, \mathrm{P}}}{R_{\star}}\bigg{)}^{2}\bigg{]}. \label{transitdepths} 
\end{equation}
Here, $F_{\lambda,\star+\mathrm{P}}$ is the combined wavelength dependent stellar and exoplanet photon flux for each exposure, $F_{\lambda,\star}$ is the wavelength dependent stellar photon flux, and $\bigg{(}\frac{R_{\lambda, \mathrm{P}}}{R_{\star}}\bigg{)}^{2}$ are the wavelength dependent transit depths calculated in Section \ref{sec:spectra} where $R_{\lambda, \mathrm{P}}$ is the wavelength dependent radius of the exoplanet and $R_{\star}$ is the stellar radius. The addition of white noise and instrumental effect is as described in \citet{dash2025detectability}, and the resultant photon flux cuboid thus obtained is denoted as \textbf{A}. These photon flux cuboids are the starting points for most HRCCS analyses in literature. 

\subsection{HRCCS analysis using \texttt{Upamana}}
\label{sec:upamana}
To detect the exoplanet signal embedded in \textbf{A}, we need to remove all sources of photon flux other than the time varying exoplanet signal. The HRCCS detrending and analysis pipeline \texttt{Upamana} \citep{dash2024constraints,dash2025detectability} achieves this by removing all sources of photon flux variations that do not vary in time (or vary at velocities lesser than the velocity resolution of the spectrograph) across wavelength by using a singular value decomposition (SVD)+multi-linear regression (MLR) based framework. This detrending procedure is ideally not supposed to remove the exoplanet signal within the first few highest ranked singular vectors because such a (weak) signal is Doppler shifted across wavelength due to the orbital motion of an exoplanet at a few km\,s$^{-1}$ (greater than the instrumental velocity resolution) throughout an observation. Thus, after detrending the flux cuboid, the exoplanet signal represents a time varying source of flux variation buried in the white noise, which can then be analysed using a cross-correlation analysis. The methodology behind the pre-processing and detrending procedures and the subsequent cross-correlation analysis utilized in \texttt{Upamana} has been described in detail in \citet{dash2024constraints} and \citet{dash2025detectability}, and we only include a brief description below for convenience to track the steps and any possible changes compared to the previous works. All analyses are assumed to be done spectral order-wise.

\begin{itemize}
    \item The low flux masking (and the threshold value utilised), standardisation, application of time-domain SVD followed by MLR (to obtain a detrended residual matrix), and subsequent injection and reprocessing of any Doppler-shifted forward template models (to obtain a reprocessed residual matrix) are performed in the exact same way as in \citet{dash2025detectability} for the case of V1298\,Tau\,b. For the case of TOI-451\,c, we use a modified criterion of 10\% of the highest 500 pixels in an order to calculate the low-flux threshold. This is because the signal for TOI-451\,c is $\sim$ 4 times weaker compared to V1298\,Tau\,b and hence requires a more thorough masking for the telluric lines.
    \item We use the CCF-to-likelihood framework from \citet{brogi2019retrieving} and \citet{gandhi2019hydra} to find the similarity between the detrended and reprocessed residual matrices. In this framework, the likelihood ($L$) values are calculated using the equation:
    \begin{equation}
        \log(L) = -\frac{N}{2}\log[s_{f}^{2} - 2SR(s) + S^{2}s_{g}^{2}], \label{logleq}
    \end{equation}
    where $s_{f}^{2}$ (data variance), $s_{g}^{2}$ (model variance) and $R(s)$ (cross-covariance) are defined as
    \begin{eqnarray}
    s_{f}^{2} & = & \frac{1}{N}\sum_{n}f^{2}(n), \nonumber \\
    s_{g}^{2} & = & \frac{1}{N}\sum_{n}g^{2}(n), \nonumber \\
    R(s)      & = & \frac{1}{N}\sum_{n}f(n)g(n-s). 
    \end{eqnarray}
    $f(n)$ and $g(n)$ are the mean subtracted values of each row from each order of the detrended residual matrix, and the corresponding row from each order from the reprocessed residual matrix. $s$ denotes a wavelength shift, corresponding to the Doppler-shift of the model template for each tested value of exoplanet orbital velocity, which is dependant on the particular exoplanet orbital solution used for injection just before the reprocessing step. $S$ is the scale factor by which the model template is multiplied to find the best fit with the data. When the model can correctly represent the data and the reprocessing step is adequate to perfectly reproduce the effect of detrending on the model template, $S =$ 1. For retrievals, we set $S$ as a free parameter (as $\log(S)$ instead). The retrieved value can inform about any biases inherent to the detrending procedure itself. Similar to the assumption in \citet{dash2025detectability}, throughout this study we set $S=1$ as from our test retrievals we find $\log(S)$ to be very close to 0 (see Figure \ref{fig:ret} for an example retrieval for V1298\,Tau\,b using the simulated nights from this study).
    \item The exoplanet orbital solution for model injection before the reprocessing step is characterised by a $v_\mathrm{sys}$-$K_\mathrm{P}$ pair (see Equations 10 to 14 in \citet{dash2025detectability} for the relevant equations), where $v_\mathrm{sys}$ is the systemic velocity of the exoplanetary system and $K_\mathrm{P}$ is the exoplanet semi-amplitude velocity. Thus, in order to detect an embedded exoplanet signal, we need to isolate the $v_\mathrm{sys}$-$K_\mathrm{P}$ pair corresponding to the injected exoplanet model from a grid of $v_\mathrm{sys}$-$K_\mathrm{P}$ pairs. For the exoplanets we use as case studies, the systemic velocity of the exoplanetary system are known from literature. Thus, we subtract out this systemic value from all $v_\mathrm{sys}$ values in the grid resulting in a $v_\mathrm{rest}$-$K_\mathrm{P}$ grid, where $v_\mathrm{rest}$ is the rest-frame velocity of the exoplanetary system. The chosen grid range and grid spacings for the analysis of each exoplanet are all described in Section \ref{sec:results}. With the chosen grids for each exoplanet, we first calculate a $\log(L)$ value for each $v_\mathrm{rest}$-$K_\mathrm{P}$ pair in the grid by using each row for each order in the detrended residual matrix and the corresponding row in the same order in reprocessed residual matrix. Then we sum across all rows in the order, and then sum across orders to calculate a single resultant $\log(L)$ value. The highest $\log(L)$ value within the $v_\mathrm{rest}$-$K_\mathrm{P}$ grid is ideally obtained at the expected exoplanet orbital solution in the case of a detection. Then we utilize the method employed in \citet{dash2024constraints} to calculate the confidence interval maps via a likelihood ratio test followed by application of Wilks' theorem \citep{wilks1938large} for 2 degrees of freedom, corresponding to a $v_\mathrm{rest}$ and $K_\mathrm{P}$ pair. A signal confidently detected appears as a series of tight, concentric contours around a certain $v_\mathrm{rest}$ and $K_\mathrm{P}$ pair, and the rest of the parameter space disfavoured by more than 4-5$\sigma$. 
\end{itemize}

\section{Results and discussion}
\label{sec:results}
\subsection{The case of V1298\,Tau\,b}
\subsubsection{Nights of observations} \label{v1298taubobs}
\begin{figure}
\centering
  \includegraphics[width=\columnwidth]{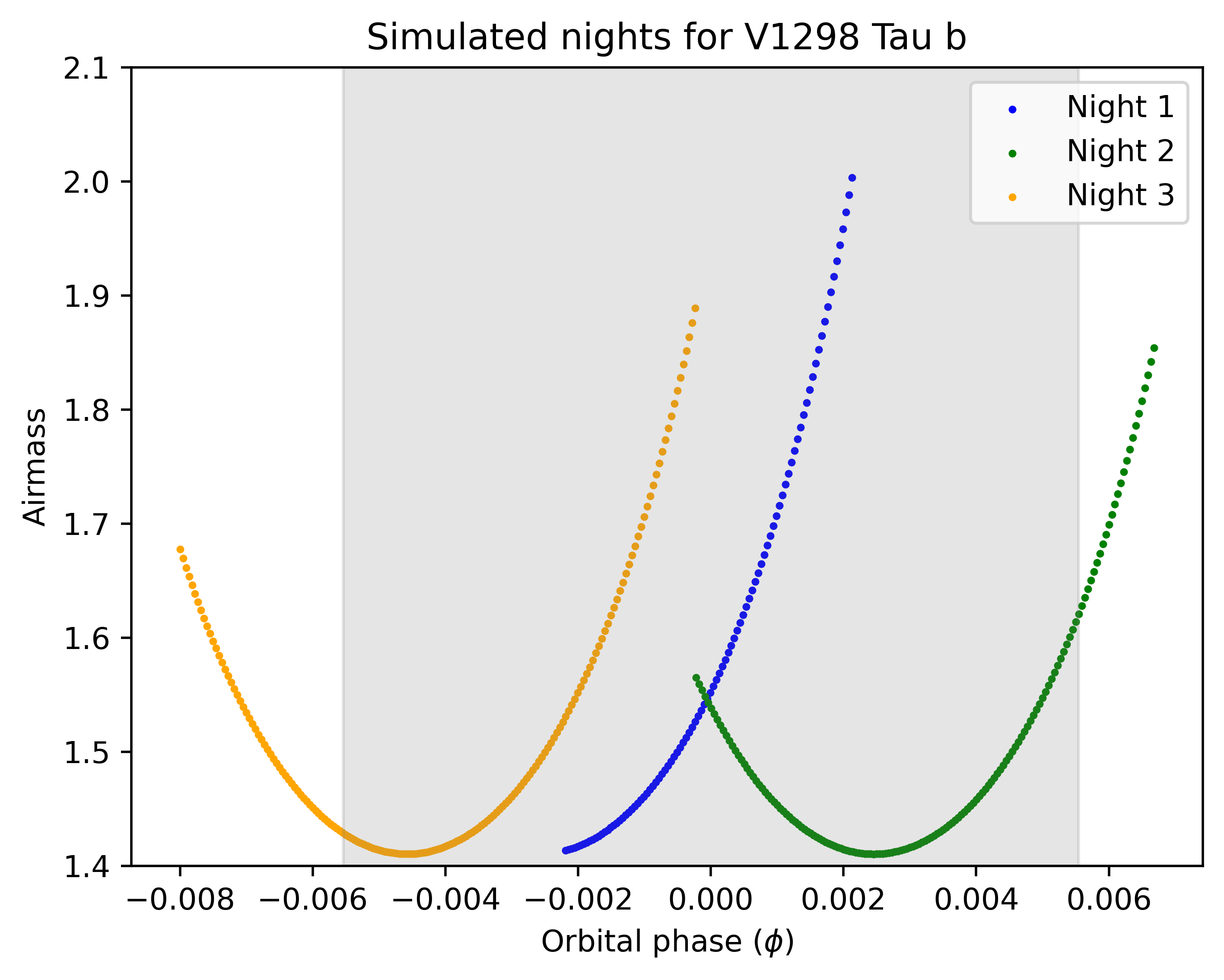}
  \includegraphics[width=\columnwidth]{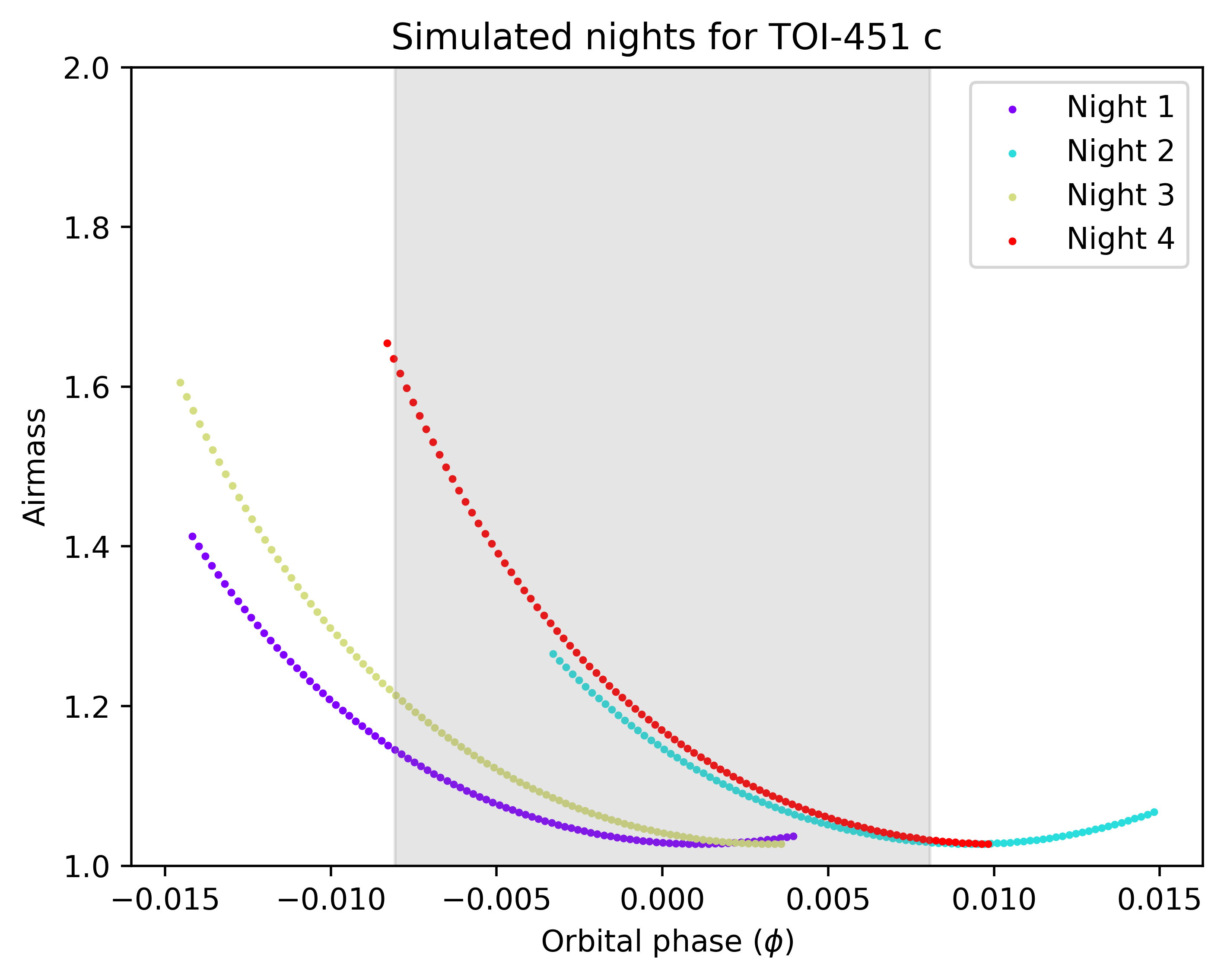}
  \caption{The variation of airmass versus exoplanet orbital phase in simulated nights using \texttt{Ratri} for the case of V1298\,Tau\,b (top panel) and TOI-451\,c (bottom panel). Phases falling inside the shaded region represent in-transit phases. For more discussions about the synthetic nights, see Sections \ref{v1298taubobs} and \ref{toi451cobs}.}
  \label{fig:v1298taub_toi451c_airm}
\end{figure}

We synthesize three nights of observations using \texttt{Ratri} showcasing the three different ways an observation of this exoplanet might be undertaken with ANDES. We depict the resulting airmass versus exoplanetary orbital phase for all three nights in Figure \ref{fig:v1298taub_toi451c_airm}. Night 1\footnote{We note that for the nights synthesized in this study, \texttt{astroplan} \citep{morris2018astroplan} used in \texttt{Ratri} will throw a warning indicating a dubious year. This is because the dates we are simulating for are far enough in the future that the IERS data used for calculation of Earth's orientation in the inherent \texttt{astropy} module are no longer valid and defaults to using a 50-year mean to do such a calculation. Thus, these dates should only be seen as indicative rather than definite.} is simulated for the night of 20th-21st January 2020, lasts 2.5\,hours (from 0100 to 0330\,hours UTC, airmass between 1.4 $\rightarrow$ 2.0) in total with all 96 exposures captured while the exoplanet is transiting its host star. Due to V1298\,Tau\,b's long transit duration, there is never a case where a complete transit for this exoplanet can be captured. The exoplanetary system is also a relatively northern target, which means that the airmass does not go below 1.4 even in the periods the target would be observable from the ELT. Night 2 is simulated for the night of 24th-25th December 2030, lasts 4 hours (from 0100 to 0500\,hours UTC, airmass ranging from 1.6 $\rightarrow$ 1.4 $\rightarrow$ 1.9) in total and has 26 of the total 153 exposures captured happening after the exoplanet has already transited its host star. Night 3 is simulated for the night of 17th-18th December 2032, lasts 4.5 hours (from 0100 to 0530\,hours UTC, airmass ranging from 1.7 $\rightarrow$ 1.4 $\rightarrow$ 1.9) in total and has 55 of the total 172 exposures captured happening before the exoplanet starts transiting its host star. For all three nights, this is close to the maximum number of exposures that can be captured from the ELT site, as in extending the limits of observation we would either reach exposures where the airmass is greater than 2 (generally the limit after which the AO system does not work optimally), or is in regions where it is either astronomical twilight or the beginning of dawn. For all three nights, the exoplanetary signal is not injected in the exposures captured at times when the exoplanet is not transiting. 
\\
\\
For all nights, we sample PWV values from a lognormal distribution with a mean of 1.25\,mm, which corresponds to the `Very Good' night parameter in \texttt{Ratri}. The mean value used is lower than the median value for PWV observed at Cerro Paranal \citep[$\sim$\,2.1\,mm, see UVES column for Table 1 in][]{kerber2012water}, the site for CRIRES+, and which is close to Cerro Armazones, which is the site for ANDES. We intentionally have this difference in median values to account for the increased elevation of the Cerro Armazones site compared to Cerro Paranal, and because we only want to model a specific way of semi-realistic lognormal time dependance of the PWVs at a given location \citep{foster2006precipitable}, rather than account for a realistic case of PWV variation from site measurements. \texttt{Ratri} uses a high-resolution PHOENIX model to approximate the spectrum of V1298\,Tau. However, this host star is a pre-main sequence star, so the PHOENIX spectrum might not be the best fit. In the absence of high-resolution models for pre-main sequence stars, we have decided to use PHOENIX for demonstration purposes. 

\subsubsection{Detectabilities} \label{v1298taubdetect}
\begin{figure*}
\centering
  \includegraphics[width=0.69\columnwidth]{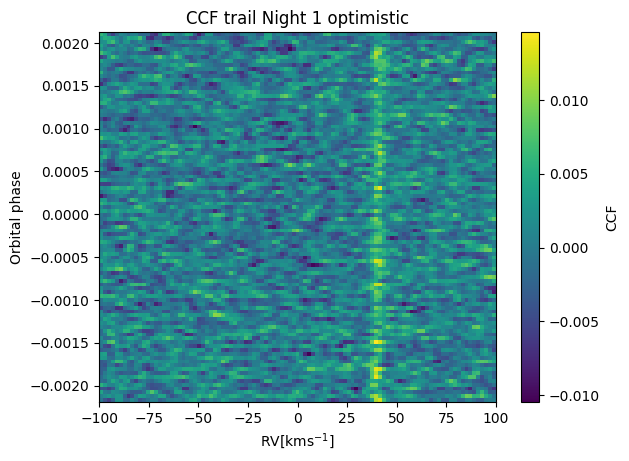}
  \includegraphics[width=0.66\columnwidth]{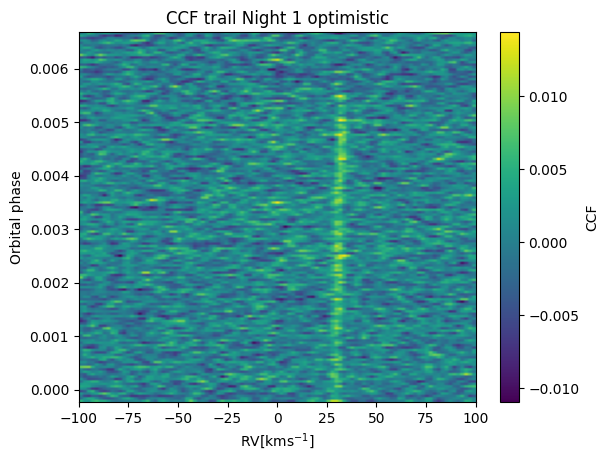}
  \includegraphics[width=0.66\columnwidth]{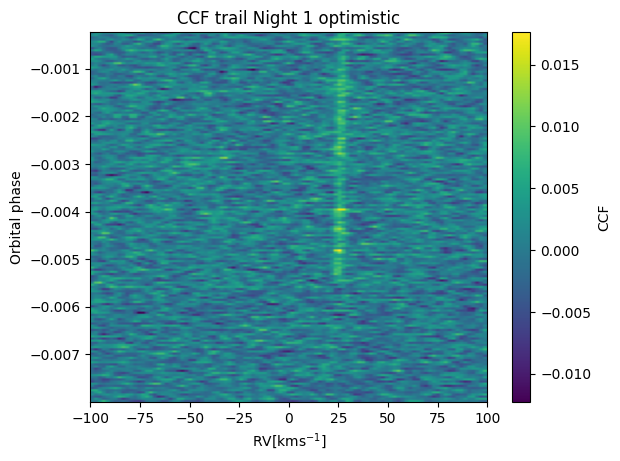}
  \includegraphics[width=0.69\columnwidth]{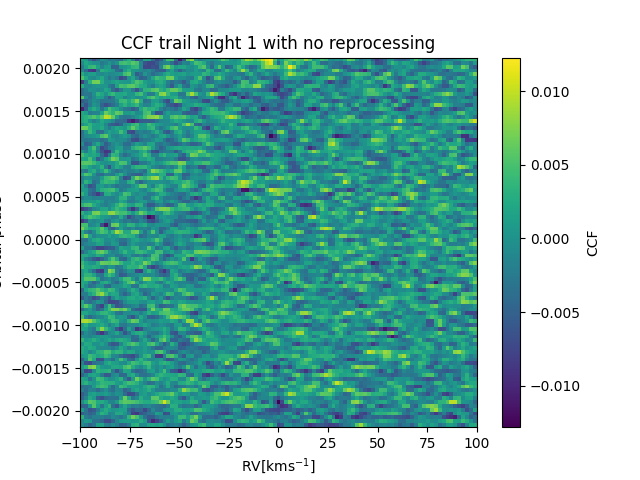}
  \includegraphics[width=0.66\columnwidth]{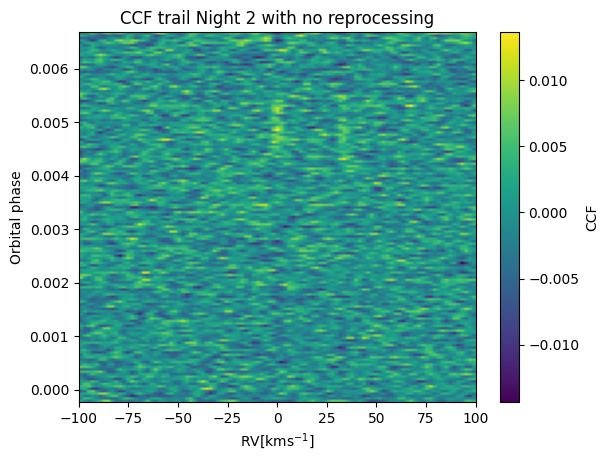}
  \includegraphics[width=0.66\columnwidth]{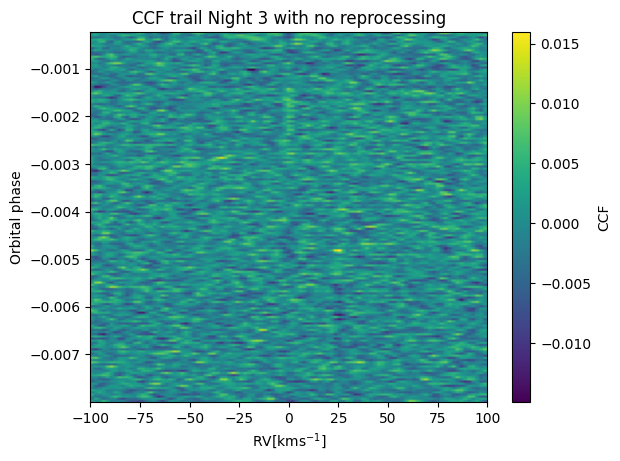}
  \caption{Exoplanet orbital trace plots for the cloud-free case of V1298\,Tau\,b for (top row) the optimistic detrending scenario and (bottom row) the non-reprocessing scenario where the SVD+MLR based detrending is followed by cross-correlation with a template model with no reprocessing performed on it. In the first row, the exoplanet orbital motion of V1298\,Tau\,b tracing the the maximum of the CCF value at each orbital phase is clearly visible in the in-transit phases for each night. However, in the non-reprocessed case the CCF trail is no longer visible for the first two nights and is only faintly visible in the third night. Additionally, the third night also shows an anti-trail of negative CCF values extending in the out-of-transit phases.}
  \label{fig:v1298taubtrails}
\end{figure*}
\begin{figure*}
\centering
  \includegraphics[width=0.67\columnwidth]{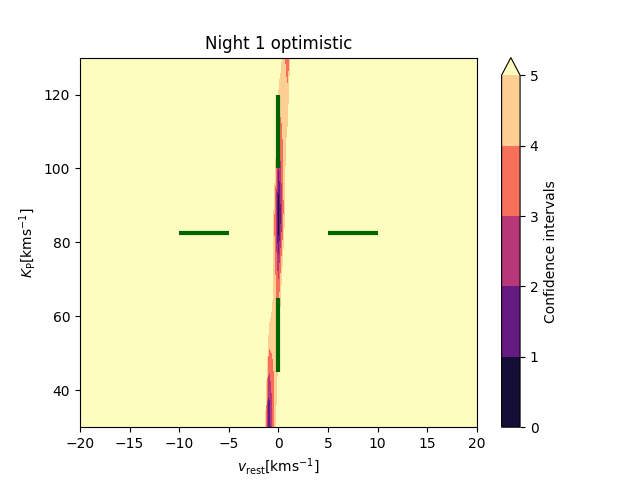}
  \includegraphics[width=0.67\columnwidth]{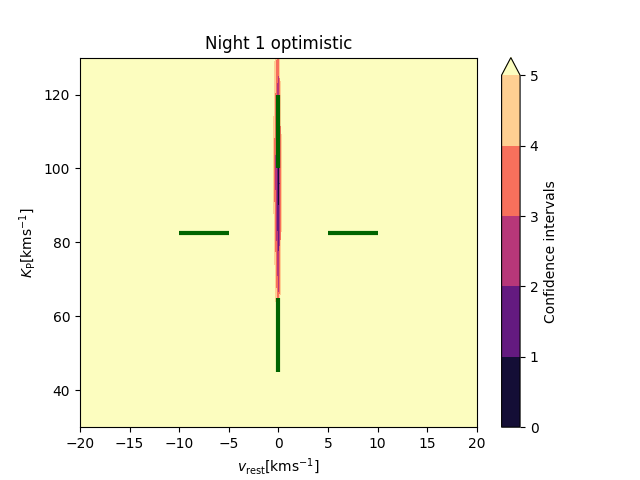}
 \includegraphics[width=0.67\columnwidth]{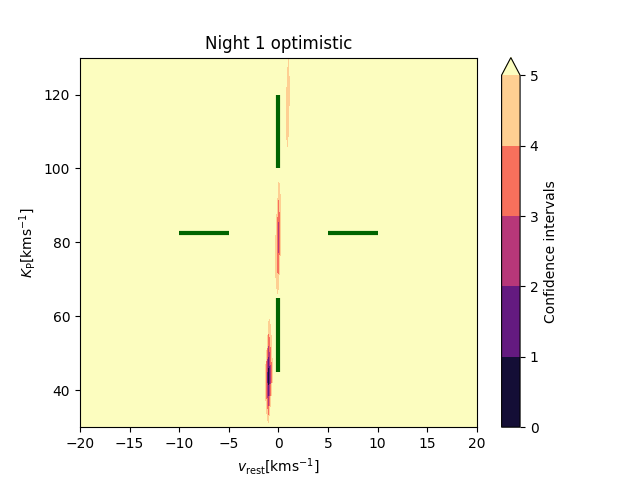}
  \includegraphics[width=0.67\columnwidth]{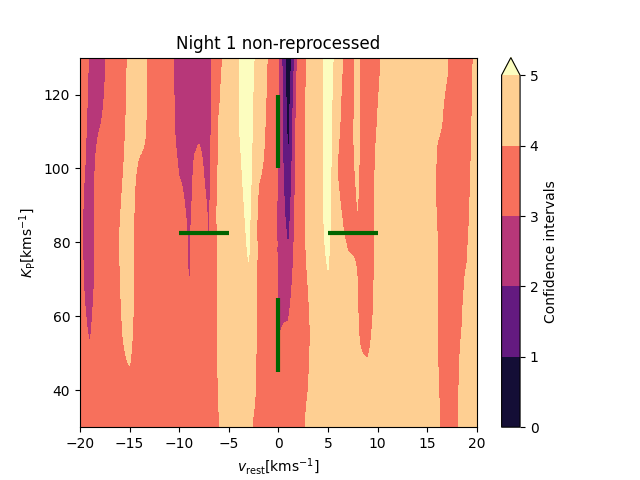}
  \includegraphics[width=0.67\columnwidth]{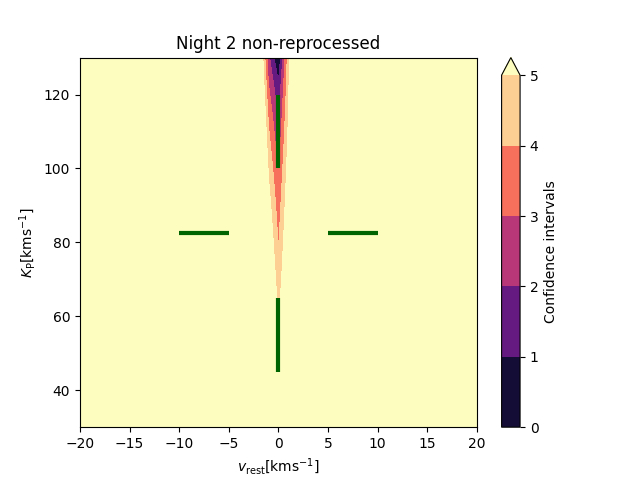}
  \includegraphics[width=0.67\columnwidth]{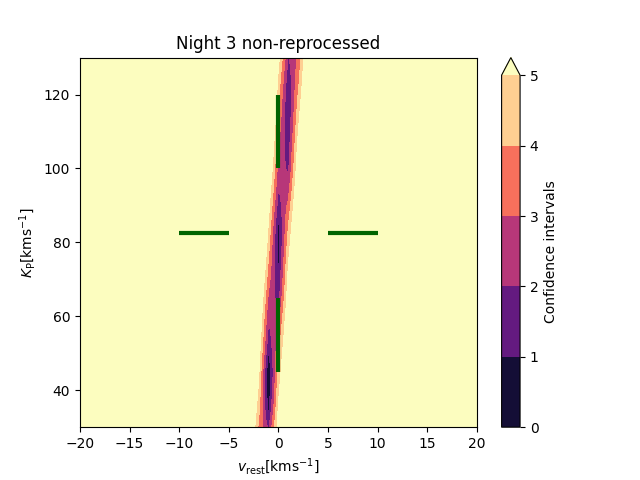}
  \caption{Recovery of an injected signal for the cloud-free case of V1298\,Tau\,b on a $v_\mathrm{rest}-K_\mathrm{P}$ grid for each synthetic night simulated using \texttt{Ratri} for the case of (top row) the optimistic detrending scenario and (bottom row) the non-reprocessing scenario where the SVD+MLR based detrending is followed by cross-correlation with a template model with no reprocessing performed on it. The injected signal is successfully detected in all three nights for the optimistic detrending scenario. For the non-reprocessing scenario, the injected signal is successfully detected only in Night 3 but with weaker constraints on $K_\mathrm{P}$ compared to the optimistic detrending case. For the other two nights, the signal is either completely removed by the detrending scenario resulting in a non-detection (Night 1) or a detection at a much higher value of $K_\mathrm{P}$ compared to the injected value (Night2). For more discussion about confidence intervals and what constitutes a detection, please see Section \ref{sec:upamana}.}
  \label{fig:v1298taubopvsnorep}
\end{figure*}

We first decided to examine if the exoplanetary signal corresponding to a cloud-free scenario for V1298\,Tau\,b injected in our datasets is detectable or not in the optimistic scenario when all the stellar flux and telluric contamination can be exactly removed. This is easy to do for our synthetic datasets as \texttt{Ratri} also produces a version of the flux cuboid with no exoplanet signal injected. Thus in the optimistic scenario, we simply divide out the synthetic flux cuboid with an exoplanetary signal with the flux cuboid with no exoplanetary signal. Then we perform the cross-correlation analysis presented in Section \ref{sec:upamana}. We present the orbital phase versus the radial velocities (RV) with which the template model is cross-correlated with the detrended data in the top row of Figure \ref{fig:v1298taubtrails}. In all three cases, the exoplanetary orbital trail tracing the maximum of the CCF values is clearly visible for all orbital values falling within the ingress and egress values. To then represent atmospheric detections on a $v_\mathrm{rest}-K_\mathrm{P}$ grid, we vary $v_\mathrm{rest}$ between -20 and 20\,kms$^{-1}$ in spacings of 1\,kms$^{-1}$, and vary $K_\mathrm{P}$ between 30 and 130\,kms$^{-1}$ in spacings of 2\,kms$^{-1}$. The grid spacings are less than the velocity resolution of the instrument but our $\log(L)$ based significance calculation is different from the significance calculated from comparison between the in-exoplanet-trail and out-of-trail CCF distributions, which are prone to an oversampling bias when the grid spacings used are less than the velocity resolution of the instrument \citep{sanchez2025robustness}. Here, we intentionally choose to slightly oversample to maximize sensitivity and avoid the case of the peak of likelihoods falling between two grid values. The $v_\mathrm{rest}$ grid is also smaller compared to studies in literature which use a S/N metric. Unlike the S/N metric which needs a sufficiently large grid to calculate an unbiased value of the standard deviation of CCFs away from the peak CCF to estimate the CCF noise, the likelihood calculation we use here does not need a larger grid since it calculates the significance by comparison of highest likelihood value versus all other likelihood values in the grid, i.e., it calculates the preference of the detected set of grid values compared to the rest of the grid. We find that we can also confidently detect the the $v_\mathrm{rest}-K_\mathrm{P}$ pair for the injected exoplanetary signal for all three nights using these grids, as shown in the top row of Figure \ref{fig:v1298taubopvsnorep}. However, the fact that the orbital trail is very straight (as shown in Figure \ref{fig:v1298taubtrails}) means that the constraints on the value of $K_\mathrm{P}$ (which represents the slope of the orbital trail) are \textbf{small} in all three scenarios. Nevertheless, the correct value lies within the 2$\sigma$ contour in all three cases. Thus, the exoplanetary signal for V1298\,Tau\,b injected in our data is at the detectable limit using HRCCS with ANDES observations.
\\
\\
Next, we look at a scenario where the detrending is performed using the SVD+MLR based algorithm in \texttt{Upamana}, but we do not reprocess our template model before we perform the cross-correlation analysis. This kind of cross-correlation analysis was used in some of the pioneering studies in the field \citep[see][as examples]{snellen2010orbital,brogi2012signature,de2013detection} to showcase detections in hot Jupiters, and are still used for HRCCS analyses today. We use a common value of $k=6$ as the number of singular vectors to be used to perform our detrending. This is an ad-hoc choice and not based on any injection-retrieval studies, but we observe that detections can be found for values extending until at least $k=10$ (no further values were tested). Additionally, we also remove the spectral orders where residual telluric contamination can be seen at RV = 0\,kms$^{-1}$ from order-wise CCF plots (corresponding to spectral orders 8, 9, 10, 11, 13 and 21 for CARMENES). The orderwise co-added orbital trace plots for this scenario are provided in the bottom row of Figure \ref{fig:v1298taubtrails}. In comparison to the corresponding optimistic detrending scenarios, no clear orbital trace is visible in the first two nights. In the third night, however, the exoplanet trail is very slightly visible at its expected location, but is then followed by an anti-trail of negative CCF values extending into the out-of-transit orbital phases. The presence of such out-of-transit anti-trails was also indicated by \citet{palle2025ground}. \citet{cheverall2024feasibility} and \citet{dash2024constraints} had pointed out that the presence of an injected signal in transmission would produce out-of-transit artefacts in the out-of transit phases after the application of a time-domain SVD based detrending procedure. The application of these artefacts is due to the inability of the detrending algorithm to properly account for the presence of an exoplanetary signal in only a few exposures, thus over-correcting for this phenomenon in the out-of-transit phases. The presence of such out-of-transit artefacts would then produce an anti-CCF trail in the our-of-transit phases if the exoplanet template is used for cross-correlation (see Appendix \ref{reprocesseffect} for a more in-depth explanation). Nevertheless, the inability to pin-point an orbital trace means that the level of detection (using only the CCF values within the ingress and egress points) varies strongly as seen in the bottom row of Figure \ref{fig:v1298taubopvsnorep}. For Night 1, there is no detection at the expected orbital parameters, with the exoplanet signal presumably being almost completely removed by the SVD based detrending scenario and only resulting in a weak signature at a far-off higher $K_\mathrm{P}$ value. Night 2 showcases a scenario with the region indicating a strong preference for the retrieved $K_\mathrm{P}$ being much higher compared to the injected value. For Night 3, there is still a detection but the constraints on $K_\mathrm{P}$ are weaker compared to its optimistic detrending scenario counterpart.
\\
\begin{figure*}
\centering
  \includegraphics[width=2.25\columnwidth]{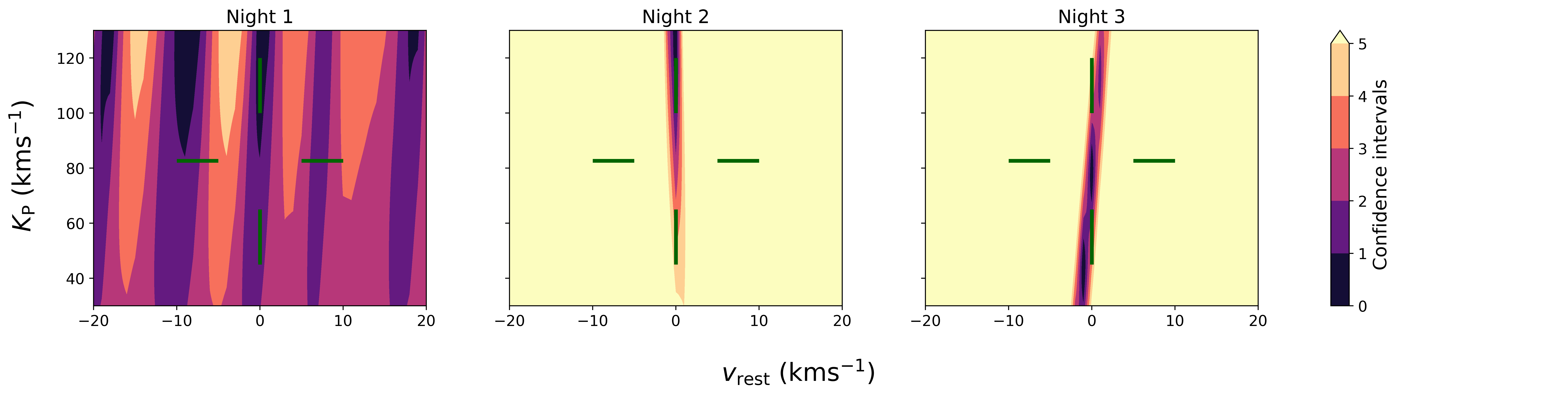}
  \caption{Recovery of an injected signal for the cloud-free case of V1298\,Tau\,b on a $v_\mathrm{rest}-K_\mathrm{P}$ grid for each synthetic night simulated using \texttt{Ratri} with the default reprocessing based application of \texttt{Upamana}. The injected signal is successfully detected in Nights 2 and 3 with the level of their detections better than the corresponding non-reprocessing scenarios in the bottom row of Figure \ref{fig:v1298taubopvsnorep}. However, we are still not able to recover the correct parameters for the injected signal in Night 1, showcasing the limit of the reprocessing procedure when no-out-of-transit spectra are present.}
  \label{fig:v1298taubdetectnights}
\end{figure*}
\\
We now look at the scenario where we run \texttt{Upamana} as usual i.e. detrending through an SVD+MLR algorithm followed by reprocessing of the template model before the cross-correlation analysis. We retain the value of $k$ used in the non-reprocessing case, as well as the values of the spectral orders not considered for the cross-correlation calculation. In this scenario, since the cross-correlation analysis (including both in-transit and out-of-transit exposures) calculating the likelihood value is calculated and stored for each value in the $v_\mathrm{rest}-K_\mathrm{P}$ grid directly, we show only the resulting detection plots in Figure \ref{fig:v1298taubdetectnights}. In comparison to their non-reprocessing counterparts, Night 2 showcases a detection at a much reduced $K_\mathrm{P}$ with the injected value falling within the 3$\sigma$ contours, and Night 3 continues to show a detection at the injected values. The fact that the detection in Night 2 is still a bit shifted together with the case that the level of detection in Night 3 is still not at the optimistic limit means that while the reprocessing step is helpful in approximating the effect of the detrending pipeline on an injected signal, it is still not perfect. Night 1 still does not showcase a detection at the expected location on the grid. Nevertheless, the application of a reprocessing step still represents the best solution to have a detection for this exoplanet conceptually, because of the effect of the detrending step on any injected signal and to also incorporate the additive effects of the trace of out-of-transit artefacts (which carry information about the injected signal, see Appendix \ref{reprocesseffect} for more details). The level of detection in Nights 2 and 3 showcases that the proportion of out-of-transit exposures also makes a difference (please see Figure \ref{fig:reprocesseffectfig2} for a visual representation of this effect). \citet{dash2024constraints} reached a similar conclusion where the presence of out-of-transit spectra led to more significant detections during injection-retrieval tests. Thus, in this scenario a higher number of out-of-transit exposures will lead to more out-of-transit artefacts, which being produced by the presence of an exoplanetary signal in the in-transit phases also carry information about the presence of an injected signal in the in-transit phases, and hence help in pin-pointing the signal better. The reprocessing step reproduces these artefacts in the out-of-transit phases, which are also different for each value in the $v_\mathrm{rest}-K_\mathrm{P}$ grid, and accounts for their cross-correlation with the out-of-transit artefacts produced in the original detrended dataset leading to a greater likelihood value for the correct parameters. Hence, Night 2 having a reduced amount of out-of-transit exposures compared to Night 3 leads to a shift in the detected value of $K_\mathrm{P}$ as there is not as much of an out-of-transit orbital anti-trace to compensate for the loss of much of the orbital trace in the in-transit phases.
\\
\begin{figure*}
\centering
  \includegraphics[width=2.3\columnwidth]{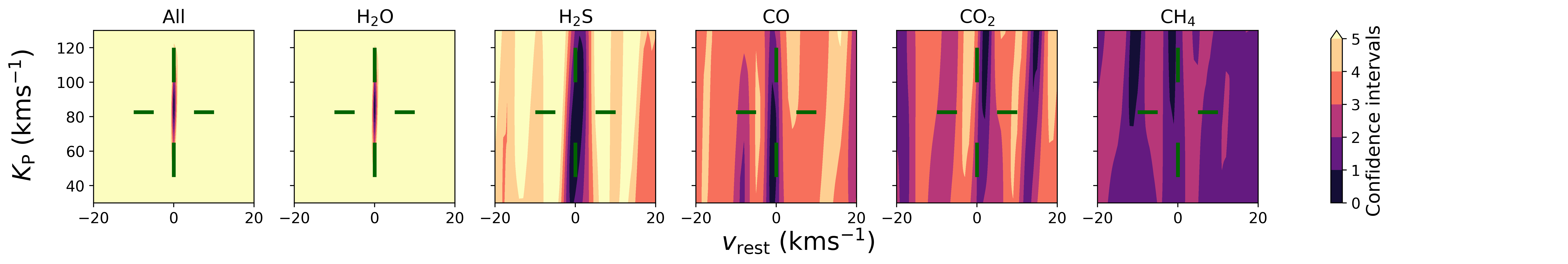}
  \includegraphics[width=2.3\columnwidth]{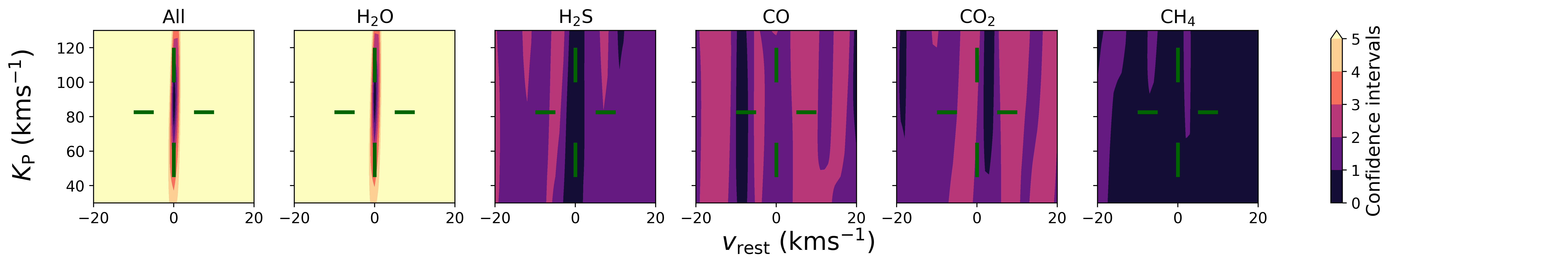}
  \caption{Recovery of an injected signal of V1298\,Tau\,b by combining the data from Nights 2 and 3 on a $v_\mathrm{rest}-K_\mathrm{P}$ grid with all molecules included (All) and only with individual molecules included in the template model (H$_2$O, H$_2$S, CO, CO$_2$ and CH$_4$) used for the cross-correlation analysis including the reprocessing step for the cloud-free case (top row) and with an opaque cloud deck placed at 0.01 bar (bottom row). The injected signal is successfully detected (All) in both cases with the level of their detections better than the corresponding individual night detections in  Figure \ref{fig:v1298taubdetectnights}. H$_2$O is also detected in both cases. H$_2$S is detected confidently only in the cloud-free case. CO is only marginally detected in the cloud-free case. Rest of the molecules are not detectable in either scenario.}
  \label{fig:v1298taubdetectsuccess}
\end{figure*}
\\
Now that we have demonstrated that Nights 2 and 3 can be used to detect the cloud-free exoplanetary spectrum for V1298\,Tau\,b individually, we broaden our cross-correlation analysis to check whether we can detect specific molecules - H$_2$O, H$_2$S, CO, CO$_2$ and CH$_4$ - individually, and compare them to the significance of the detection when all molecules are included to calculate the template spectrum. Individual template spectra are calculated by repeating the procedure in Section \ref{sec:spectra} but then including only the opacity of the molecule, in addition to CIA absorption and Rayleigh scattering due to the presence of H$_2$ and He. This is equivalent to the molecule-wise analysis shown in \citet{parker2025limits}, and is representative of analysis done using real datasets in literature where the real exoplanetary signal has all molecules included, but the preliminary search is conducted for each molecule of interest. The results for this analysis is shown in the top row of Figure \ref{fig:v1298taubdetectsuccess} on the same $v_\mathrm{rest}-K_\mathrm{P}$ grid used for the reprocessing task for each night individually. However, we have now combined the results from both nights by adding the likelihoods calculated from HRCCS analysis from each night, and then calculating the significances. As expected, combining two nights of data results in a much stronger detection and tighter constraints on the values for $v_\mathrm{rest}$ and most noticeably $K_\mathrm{P}$ for the template spectrum with all molecules included, compared to each individual night in Figure \ref{fig:v1298taubdetectnights}. H$_2$O being an abundant molecule in the atmosphere of this exoplanet, and with a significant number of lines in the YJH band, also results in a strong detection, with the same constraints on the orbital parameters as the case with all molecules included in the template spectrum. H$_2$S also shows a strong detection, but with much weaker constraints on $K_\mathrm{P}$. CO presents a weaker \textbf{(marginal)} detection compared to the other molecules as most of the parameter space is excluded at $\sim$4-5$\sigma$ compared to $>$5$\sigma$ for H$_2$O and H$_2$S. This is expected as CO only has lines in the H-band, which represents a small fraction of the total wavelength coverage. CO has more lines in the K-band, and considering that it is already marginally detectable using the YJH bands of ANDES, it should be readily detectable if a planned K-band incorporation into ANDES moves ahead in the future. CO$_2$ and CH$_4$ show no detections, showcasing that even though they have a number of lines in the YJH band, their line depths are far too weak compared to the level of noise in the datasets to stand-out during the HRCCS analysis, even with two nights of observations. We note that since we used the order-wise wavelength range of the CARMENES CCD array to construct our synthetic datasets, the detections shown here represent a lower limit of the final significance achieved that can be achieved in reality as there is a difference in 0.1\,$\mu$m in the wavelength coverage of the H band between CARMENES (until 1.7\,$\mu$m) and ANDES (until 1.8\,$\mu$m).
\\
\\
To compare with the cloud-free results obtained above, we repeat the same analysis but now with a cloud-deck model added at the level (0.01 bar) showcased by JWST results in \citet{barat2025metal}. The results are shown in the bottom panel in Figure \ref{fig:v1298taubdetectsuccess}. There is an overall decrease in the final significance and weaker constraints on $K_\mathrm{P}$ compared to the cloud-free scenario for all detected molecular species, as well as the case with all molecules included in the template model. While the complete model and H$_2$O alone would still be easily detectable, H$_2$S and CO now are no longer detectable, compared to their cloud-free counterparts, because most of the parameter space is now excluded only at $\sim$2$\sigma$. 
Thus, when using a complete Bayesian retrieval, similar to \citet{gibson2022relative}, we expect strong constraints on H$_2$O and the orbital parameters for V1298\,Tau\,b, but optimistically only upper limits on the abundances for the rest of the species. Due to time and resource constraints, that kind of retrieval is outside the scope of this work, which instead only focusses on detectability of molecular species.

\subsection{The case of TOI-451\,c} 
\label{451night}
We use the findings from our case study of V1298\,Tau\,b to check whether TOI-451\,c, which has a much lower orbital period of $\sim$9.19\,days, and a transit time of 3.56\,hours is also detectable using ELT-ANDES.

\subsubsection{Night of observation} 
\label{toi451cobs}
TOI-451\,c is a southern target and would hence be observable easily from Cerro Armazones. However, the host star is comparatively further away compared to V1298\,Tau, so we found\footnote{using a time-invariant and optimistic scenario with the ratri\_1D.py script} and subsequently use an exposure time of 120\,s to have a similar SNR /resolution element in each exposure as in the case of V1298\,Tau. We use the automatic planning and simulating tool in \texttt{Ratri} to simulate 4 nights of `Very Good' (thus a mean PWV of 1.25\,mm for the lognormal sampling) observations using ANDES : a 4.2\,hour (from 0500 to 0900\,hours UTC) observation (Night 1) for the night of 27th and 28th September 2028 with the airmass variation between 1.03 and 1.41, a 4.2\,hour (from 0450 to 0850\,hours UTC) observation (Night 2) for the night of 09th and 10th October 2029 with the airmass variation between 1.03 and 1.26, a 4.2\,hour (from 0520 to 0920\,hours UTC) observation (Night 3) for the night of 14th and 15th September 2030 with the airmass variation between 1.03 and 1.6, and a 4.2\,hour (from 0040 to 0440\,hours UTC) observation (Night 4) for the night of 21st and 22nd November 2032 with the airmass variation between 1.03 and 1.65. These nights with the airmass variation throughout each night are depicted in the bottom panel of Figure \ref{fig:v1298taub_toi451c_airm}. We have a high proportion of out-of-transit exposures for Nights 1-3 since our previous case study shows that the reprocessing procedure of \texttt{Upamana} works best in such a case. Night 4 has fewer out-of-transit exposures, but has them divided between ingress and after egress. 

\subsubsection{Detectabilities and differentiability} \label{toi451detectdiff}
\begin{figure*}
\centering
  \includegraphics[width=2.25\columnwidth]{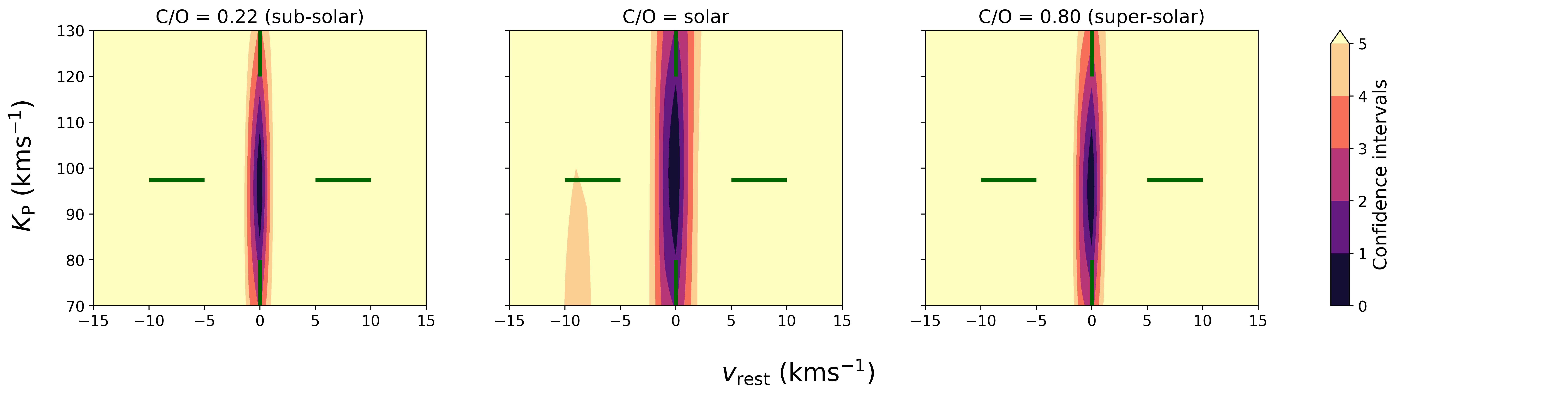}
  \includegraphics[width=2.25\columnwidth]{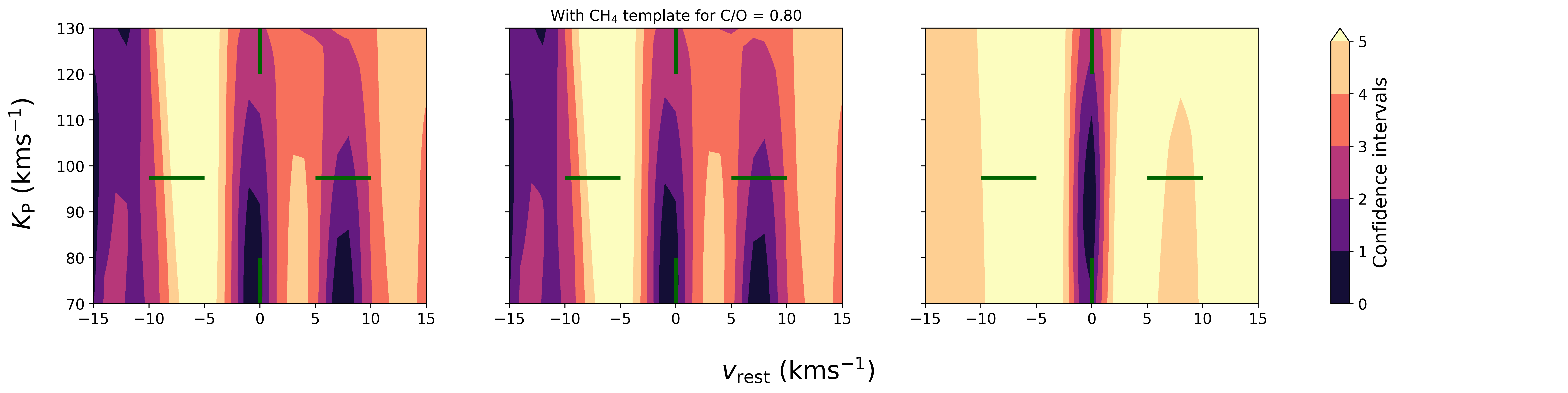}
  \caption{(Top row) Recovery of an injected signal of TOI-451\,c corresponding to differing C/O ratios of 0.22 (sub-solar), 0.55 (solar) and 0.80 (super-solar) on a $v_\mathrm{rest}-K_\mathrm{P}$ grid for Night 1 (see bottom panel of Figure \ref{fig:v1298taub_toi451c_airm}) simulated using \texttt{Ratri} with the default reprocessing based application of \texttt{Upamana}. All cases are detectable, but the solar case has weaker constraints on the recovered value of $K_\mathrm{P}$. (Bottom row) Same analysis as the previous row, but with a template model of CH$_4$ obtained using abundances and P-T values from the C/O$=$0.80 case. CH$_4$ is detectable in the super-solar case, but presents pseudo-structures due to aliasing effects in the other two cases.}
  \label{fig:toi451cdetectnights}
\end{figure*}
We first examine the detectability of TOI-451\,c by assuming three different values for the atmospheric C/O ratio, synthesizing the night of observation using \texttt{Ratri} as in Section \ref{451night} with the corresponding template model injected into the dataset, and then running \texttt{Upamana} with each of these template models to perform the cross-correlation analysis. We then check if all three cases would be differentiable from each other. For the detectability analysis, we run \texttt{Upamana} in its default mode with the reprocessing step included. This time we use $k=8$ and remove the same spectral orders as in the case of V1298\,Tau\,b from our cross-correlation analysis. Compared to the previous case study, we vary $v_\mathrm{rest}$ between -15 and 15\,kms$^{-1}$ in intervals of 1\,kms$^{-1}$, and vary $K_\mathrm{P}$ between 70 and 130\,kms$^{-1}$ in intervals of 2\,kms$^{-1}$. The results with the combined likelihoods from all four simulated nights are shown in the upper row of Figure \ref{fig:toi451cdetectnights}. It is easily seen that both the sub-solar and super-solar cases are well detectable, followed by the solar case which has weaker constraints on $K_\mathrm{P}$. From Figure \ref{fig:spectra}, the line depths for the sub-solar case are seen to be the deepest, followed by the solar and super-solar cases. This explains the easier detectability of the sub-solar case. Even though the super-solar case also has very similar line depths as the solar case, the increase in CH$_4$ abundance leads to an increase in the number of lines to now include H$_2$O, CO and CH$_4$ in comparison to mostly H$_2$O and CO dominated in the sub-solar and solar cases. This leads to an increase in the significance of detection for the super-solar case over the solar case as more lines are included in the cross-correlation process.
\\
\begin{figure}
\centering
  \includegraphics[width=1\columnwidth]{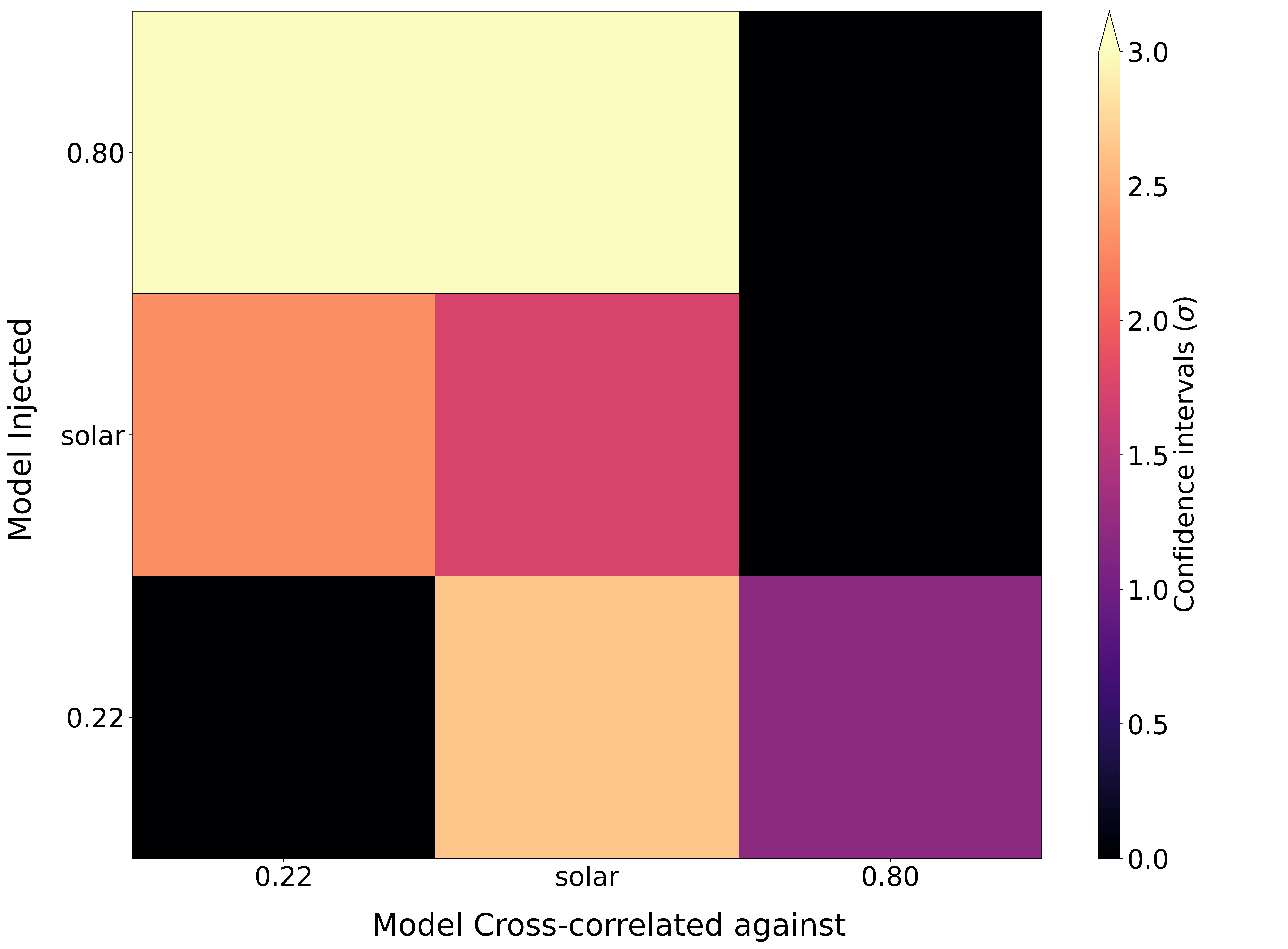}
  \caption{Grid-based model selection plot for the case of TOI-451\,c. 4 nights of synthetic data are enough to differentiate between the super-solar and the rest of the C/O cases (sub-solar and solar) by rejection of the models away from C/O $=$ 0.8 at . However, the sub-solar and solar C/O cases cannot be precisely differentiated using the same nights, with the solar case also potentially favouring the higher C/O case. See Section \ref{toi451detectdiff} for further discussion.}
  \label{fig:toi451cdiff_1_night}
\end{figure}
\\
Now we check if the different C/O cases would be differentiable from each other. We do this in two ways. For the first method, we take advantage of the fact that CH$_4$ would be a major molecule of interest in the super-solar case and repeat the previous detectability analysis, but with a template of CH$_4$ constructed with abundance and P-T values from super-solar case. The results for this analysis is shown in the bottom row of Figure \ref{fig:toi451cdetectnights}, and shows that it is only in the super-solar case where CH$_4$ is comfortably detected (as expected). For the other two C/O cases, while CH$_4$ is not detected, there is aliasing through cross-correlation with the background noise to produce pseudo-structures across the grid, but notable close to the expected injected signal values that it might become tantalising to think of these are possible weak signatures, even when they are not. It is only in the super-solar C/O case that the detection becomes prominent enough to rise above these pseudo-structures. This analysis thus provides a cautionary tale about over-interpreting structures produced across a detection grid.
\\
\\
For the second method, we utilize the same approach applied for model differentiability in \citet{dash2025detectability}. We first calculate three $\log(L)$ grids, one each from cross-correlating all synthetic data cuboids with each template model, but on a much reduced $v_\mathrm{rest}-K_\mathrm{P}$ grid, with $v_\mathrm{rest}$ ranging between -5 and 5\,kms$^{-1}$ in 1\,kms$^{-1}$ intervals and $K_\mathrm{P}$ ranging between 80 and 120\,kms$^{-1}$ in intervals of 2\,kms$^{-1}$ (with the expected exoplanet $K_\mathrm{P}$ value falling roughly in the middle of this range). The maximum likelihood obtained within this grid is saved as the corresponding $\log(L)$ value for the cross-correlation of a specific template model with the constructed flux cuboid. At the end, we are left with a $3\times3$ matrix of $\log(L)$ values and we calculate the significance values for each row of this matrix as outlined at the end of Section \ref{sec:upamana}, but now with 3 degrees of freedom. The choice of three degrees of freedom is motivated by the fact that we only changed the amount of oxygen in the atmospheric chemistry simulations as a proxy to change C/O ratios, while leaving everything else the same. So, there is only an additional free factor of oxygen abundance apart from the usual two degrees of freedom representing $v_\mathrm{rest}$ and $K_\mathrm{P}$. At the end of the analysis, we are left with with a model selection on a grid plot as shown in the left panel of Figure \ref{fig:toi451cdiff_1_night}. If each injected model would be perfectly selected in each row through the cross-correlation analysis, we would be left with an auto-correlation diagonal with the other two models rejected. However, the model being selected is correct only in the sub-solar and super-solar C/O cases, which means that the diagonal is not exactly retrieved. The selection is the strongest for the super-solar C/O case where the other scenarios are excluded at $>$3$\sigma$. This is because CH$_4$ can be detected strongly in the super-solar case as we found in the previous analysis, which tilts the selection heavily in favour of auto-correlation. For the sub-solar case, while the preference for auto-correlation is correct, the super-solar case is still not rejected. Similarly, the solar case also shows a wrong preference for the super-solar case, in comparison to its own auto-correlation. These two results might be explained by the aliasing effect we showed in the previous CH$_4$ detectability analysis. While in the sub-solar case, the injected signal is strong enough to offset the effect of aliasing, but still not exclude the super-solar C/O case which has CH$_4$ lines, the solar case is not able to do the same because of the much weaker line strengths of its H$_2$O lines coupled with the presence of a few CH$_4$ lines. Thus, using our four simulated nights, it is only possible to say confidently whether the atmosphere of this exoplanet would be super-solar when it is indeed super-solar, but not whether it is definitively sub-solar or just solar, when the atmosphere is either of these two cases. Thus, we would need a few more nights (or even require the planned K-band incorporation in ANDES to go ahead) to place precise constraints on the C/O of the planetary atmosphere if the atmosphere of TOI-451\,c is sub to solar in composition, but the number of nights (i.e. 16.8\,hours of observations) is already enough to confidently place precise constraints on the C/O if the atmosphere is super-solar. Since this work is primarily focussed on detectability of molecular species in our exoplanets of interest, we do not proceed to make a detailed analysis regarding characterisation of C/O ratios using more nights or using retrievals.

\section{Conclusions}
\label{sec:conclusion}
In this study, we investigated the feasibility of detecting and characterizing the atmospheres of the young, long-period sub-Neptunes V1298\,Tau\,b and TOI-451\,c at high spectral resolution with ELT/ANDES. We constructed atmospheric models informed by HST+Spitzer+JWST constraints for V1298\,Tau\,b and explored a range of C/O ratios for TOI-451\,c. High-resolution transmission templates were generated and injected into synthetic YJH-band observations produced with the \texttt{Ratri} pipeline. These datasets were then analyzed using an HRCCS detrending and cross-correlation framework (SVD+MLR-based) to evaluate the recovery of the injected Doppler-shifted signals and their corresponding orbital parameters. Our main conclusions are summarized below.
\begin{itemize}
    \setlength\itemsep{0.5em}
    \item We synthesize 3 nights of observations, each lasting between 2.5-4\,hours, for V1298\,Tau\,b. All three nights should have resulted in a tight detection of the exoplanetary atmosphere in the optimistic scenario, but the actual level of detection varies across the nights. 
    \item In the night with no out-of-transit exposures, we fail to detect the exoplanetary signal whether we reprocess the template model before the cross-correlation step or not. Thus, this kind of observation represents a scenario where we expect the exoplanet signal to be almost completely eroded away by the detrending procedure. We propose that such observations be avoided for detecting and characterising long-period exoplanets. Additionally, this also motivates future studies of better and more non-invasive ways of detrending high-resolution datasets as the current SVD/SYSREM based approaches do not let us utilise the full potential of having a larger telescope, especially for long-period exoplanets. Reprocessing is a time and resource intensive process as well, so it is worthwhile to consider alternative detrending approaches which can remove telluric contamination while not affecting the embedded exoplanet signal. Direct and more accurate modeling of telluric lines (similar to \citet{lockwood2014near}) and machine-learning approaches like one proposed in \citet{meech2022applications} represent potentially attractive avenues.
    \item The other two nights have a significant number of out-of-transit exposures and result consistently in detections when the template model is reprocessed before cross-correlation. If the model is not reprocessed, either the obtained orbital parameters are far off from the injected values, or present less significant detections at the injected parameters. Thus, the reprocessing step is essential in HRCCS analysis with SVD/SYSREM based detrending even if the only objective is to look at predictions of detection levels across a grid of spectral templates. Failure to perform this step in such analyses might result in underestimating the overall level of detectability. Similarly, studies assuming an optimistic/perfect level of detrending for detectability and characterisation studies might overestimate the actual detectabilities. We note, however, that these results are based on a robust CCF-to-$\log(L)$ framework for calculation of significances, compared to the more widespread S/N metric or the metric for calculating detection significances based on a comparison of in-exoplanet-trail and out-of-trail distributions.
    \item The fact that the presence of out-of-transit exposures increases the significance of detection was investigated and proposed already in \citet{cheverall2024feasibility}, \citet{dash2024constraints} and \citet{palle2025exploring}. Our findings are consistent with and augments the results of these studies and also shows the drastic scenario for long-period exoplanets where the absence of out-of-transit exposures can result in non-detections, even when the exoplanet is optimistically well within the limit of detectability with the instrument. Thus, we reiterate that some amount of out-of-transit exposures be taken during observations of these exoplanets. While this study is restricted to transmission, a similar phenomenon of the exoplanet signal being present in some phases and absent in other can occur for emission spectra when the exoplanet is occulted by the star. Thus, a similar phenomenon of including in-occultation phases and out-of-occultation phases together during detrending is also worth considering (and also proposed in \citet{cheverall2024feasibility}), but a detailed analysis is beyond the scope of this study.
    \item For the case of V1298\,Tau\,b, with the best fit template spectrum constructed based on the atmospheric parameters found to fit for HST+Spitzer and JWST observations in \citet{barat2025metal}, while the exoplanetary atmosphere itself is easily detectable with the YJH band of ANDES, individual molecules themselves are detected at varying significances. With two nights of synthetic observations lasting 8.5\,hours in total (in which the exoplanet is detectable), we expect H$_2$O and H$_2$S to be detectable, and CO to be marginally detectable. The detectability of these molecules is highest in the cloud-free case, which is expected as this case presents the deepest line depths (see Figure \ref{fig:spectra}). A cloud deck case as suggested from \citet{barat2025metal} is still detectable, but only H$_2$O is easily detectable in the same time-frame. The rest of molecules detected in \citet{barat2025metal} are either not detectable in the YJH band or would require much longer observing time.
    \item TOI-451\,c is also detectable with four 4.2\,hours (i.e. 16.8\,hours in total) of observation in the YJH band of ANDES, but the level of detectability varies slightly between the C/O ratios assumed to model the atmospheric chemistry. The sub-solar and super-solar C/O ratio cases are more strongly detectable, with the solar C/O case placing slightly weaker constraints on the $K_{P}$ in comparison to the other two cases. The observations are enough to strongly differentiate between the super-solar case and the rest, but not the sub- and solar cases from the other cases. While the sub-solar case tends to prefer the correct C/O case during model selection, the solar case erroneously prefers the super-solar case due to aliasing effects. Thus, more observations are needed to increase the fidelity of differentiation for the sub- and solar cases.
    \item A comparative analysis of the modelled atmospheric chemistries of V1298\,Tau\,b and TOI-451\,c demonstrates that young sub-Neptunes with similar bulk C/O ratios can nevertheless exhibit different molecular compositions. In particular, sulfur-bearing species such as SO$_2$, which is abundant in the atmosphere of V1298\,Tau\,b, are strongly suppressed in TOI-451\,c even for the same sub-solar C/O ratio. These differences can be attributed to a combination of factors, including the contrasting ultraviolet radiation fields and spectral resolutions of the adopted stellar UV spectra, differences in assumed bulk metallicity, and variations in orbital separation and atmospheric structure. This highlights that elemental ratios alone are insufficient to fully characterise young sub-Neptune atmospheres and underscores the importance of combining JWST constraints with high-resolution ELT observations to disentangle the physical drivers of atmospheric chemistry.
    \item The results of this study thus suggest that not only will the HRCCS technique continue to be viable for young, long-period sub-Neptune progenitors and sub-Neptunes themselves in the ELT era, it will also be an essential tool in characterising the diversity of this population. However, as can be seen from Appendix \ref{reprocesseffect}, it is necessary to select for a proper in to out-of-transit spectral ratio in order to maximise the exoplanet trail composed of the highly attenuated in-transit trail and the attenuated but still prominent out-of-transit anti-trail from out-of-transit artefacts. The long-periodicity of the results obtained here are encouraging for the case of more terrestrial exoplanets in similar or even longer period orbits as well. The detectabilities of such exoplanets need to be similarly examined in future studies, but they are going to be very challenging considering that their signatures are even weaker. While detections might still be possible by considering many nights of observation, precise characterisation would be difficult without a very high amount of observed nights to offset aliasing effects similar to those seen in the solar C/O case of TOI-451\,c in this study.
\end{itemize}

\begin{acknowledgments}
SD acknowledges support through a Postdoctoral Fellowship from Université Paris Cité. LM acknowledges funding support from the DAE through the NISER project RNI 4011.
\end{acknowledgments}

\begin{contribution}
DD and LM were involved in initial planning of the project and then brought SD into the fold. SD, DD and LM together then together formulated a definite plan and work schedule, followed by SD and DD performing the high-resolution analysis and chemistry modelling respectively based on multiple subsequent deliberations. SD and DD were together responsible for writing the manuscript. All authors were then equally responsible for editing and correcting the manuscript.  


\end{contribution}

%
\facilities{Simulated observations using ELT/ANDES}

\software{astropy \citep{astropy:2013,astropy:2018,astropy:2022}, numpy \citep{harris2020array}, scipy \citep{virtanen2020scipy}, astroplan \citep{morris2018astroplan}, Ratri \citep{dash2025detectability}, Upamana \citep{dash2024constraints,dash2025detectability}}


\appendix

\section{Comparison of \texttt{Ratri} with the ANDES exposure time calculator} \label{andesratricomp}

\begin{figure*}[h!]
\centering
  \includegraphics[width=\columnwidth]{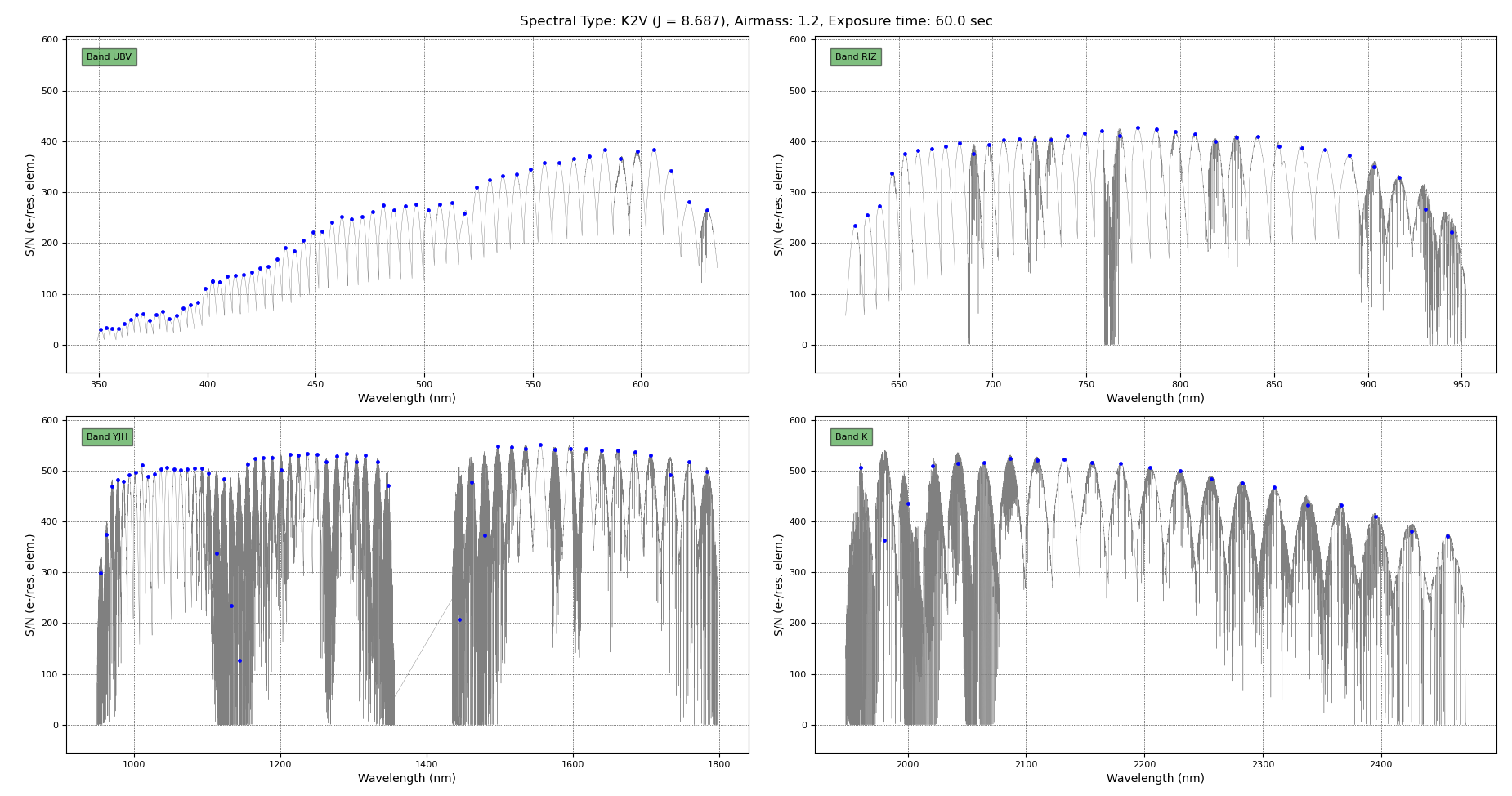}
  \caption{The SNR/resolution element across all possible bands for a V1298\,Tau like star obtained using the ANDES-ETC simulator \citep{sanna2024andes,palle2025exploring}. The exposure time is 60\,s and the airmass value assumed to model the transmission of Earth's atmosphere is 1.2.}
  \label{fig:ANDESexp}
\end{figure*}

\begin{figure*}
\centering
  \includegraphics[width=0.45\columnwidth]{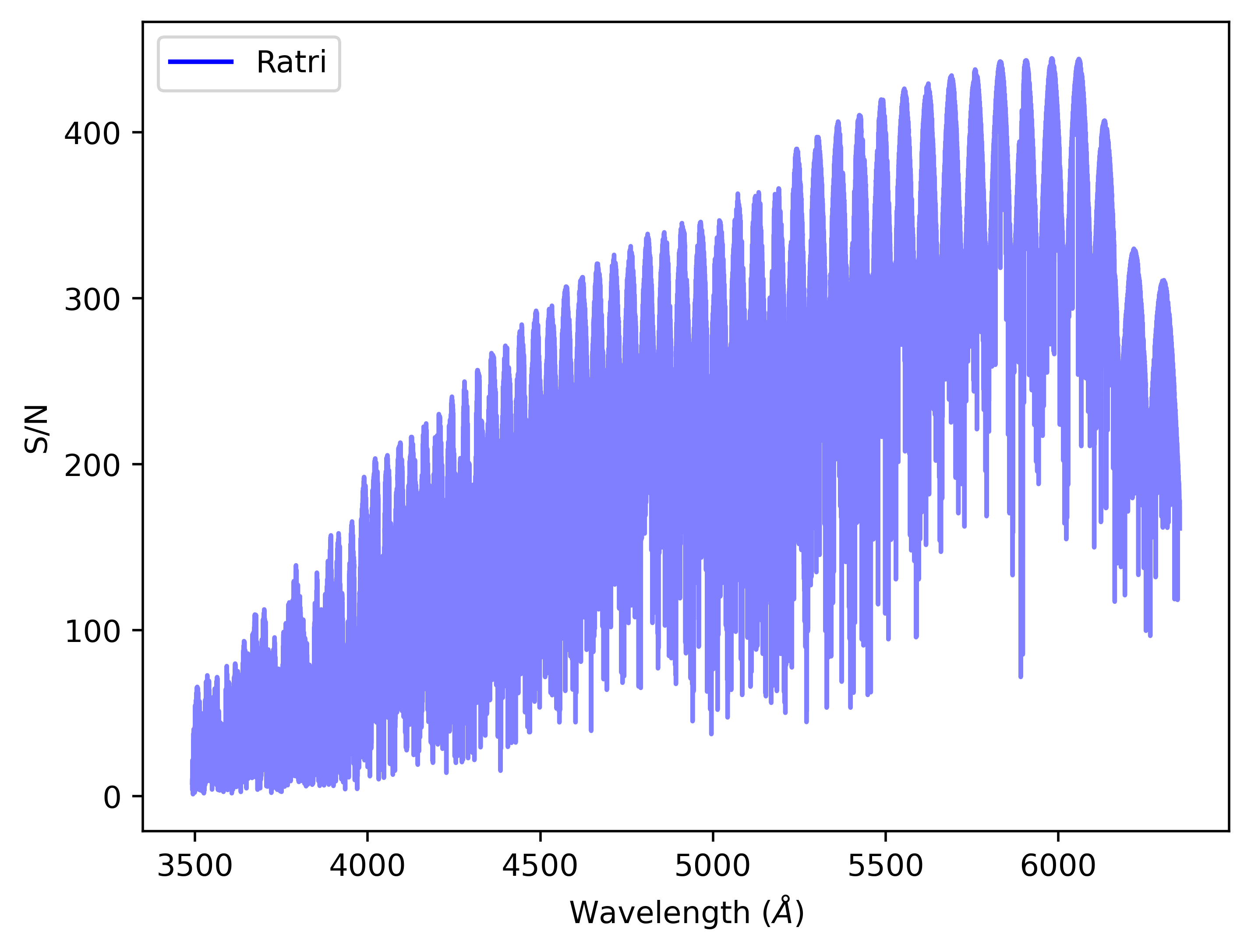}
  \includegraphics[width=0.45\columnwidth]{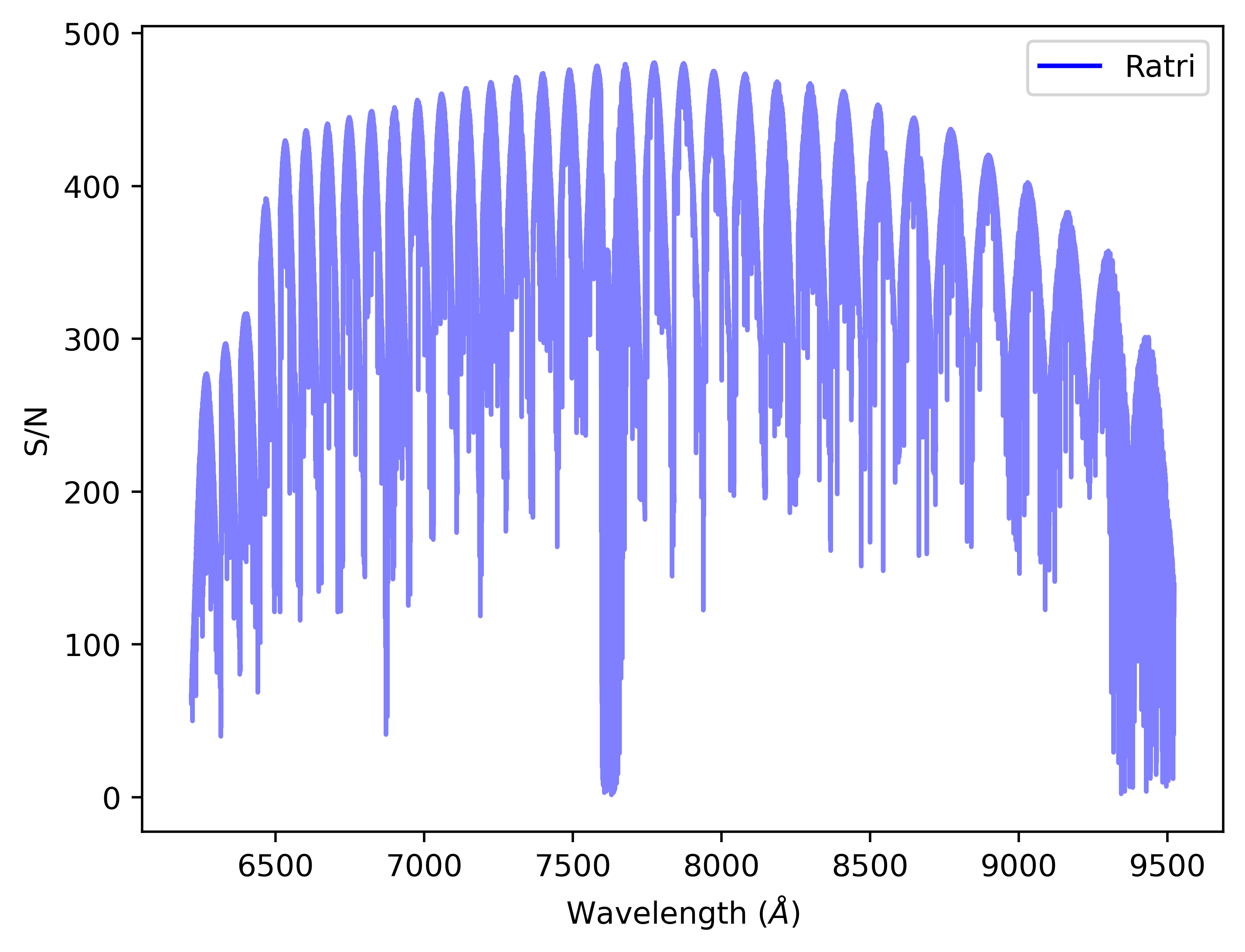}
  \includegraphics[width=0.45\columnwidth]{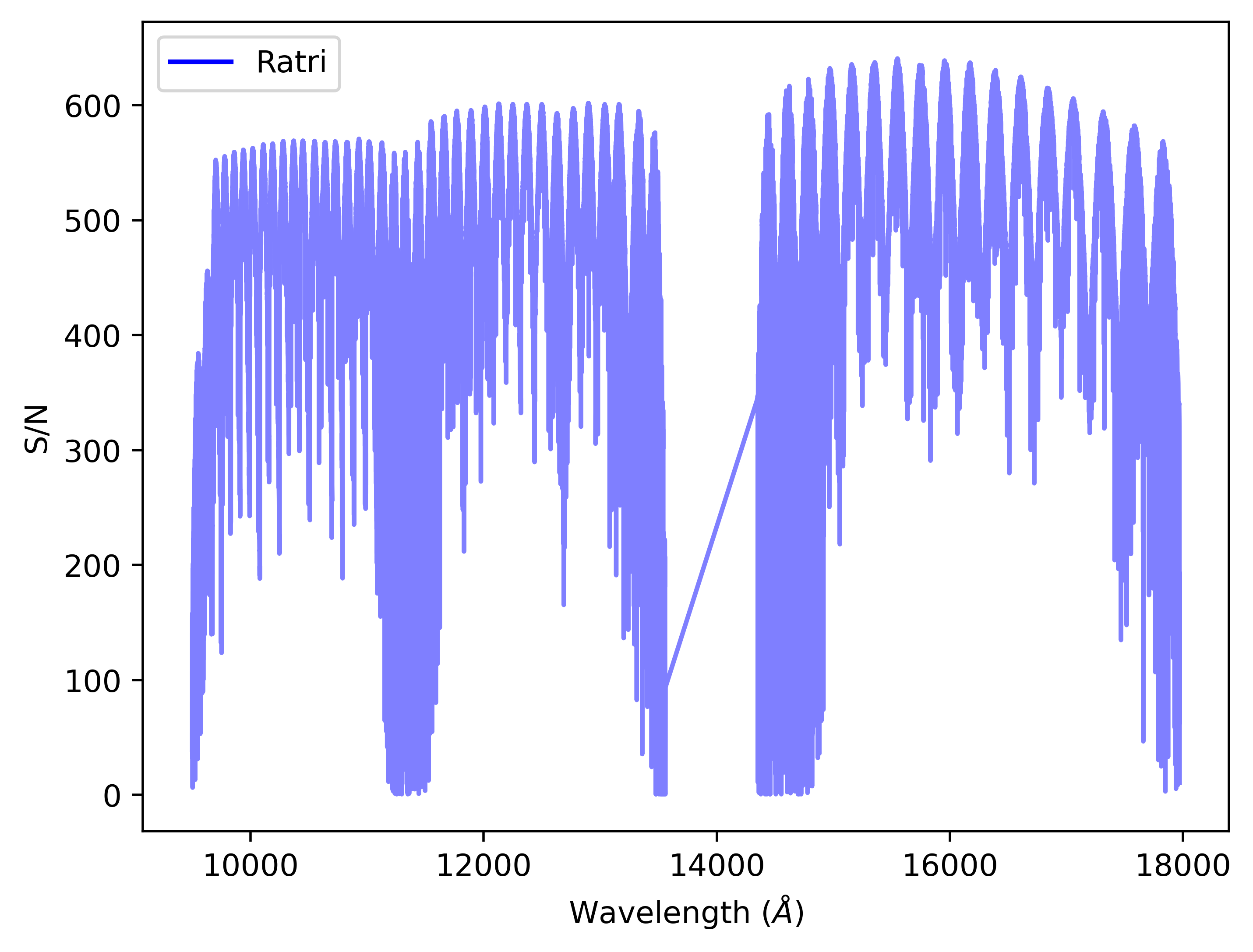}
  \includegraphics[width=0.45\columnwidth]{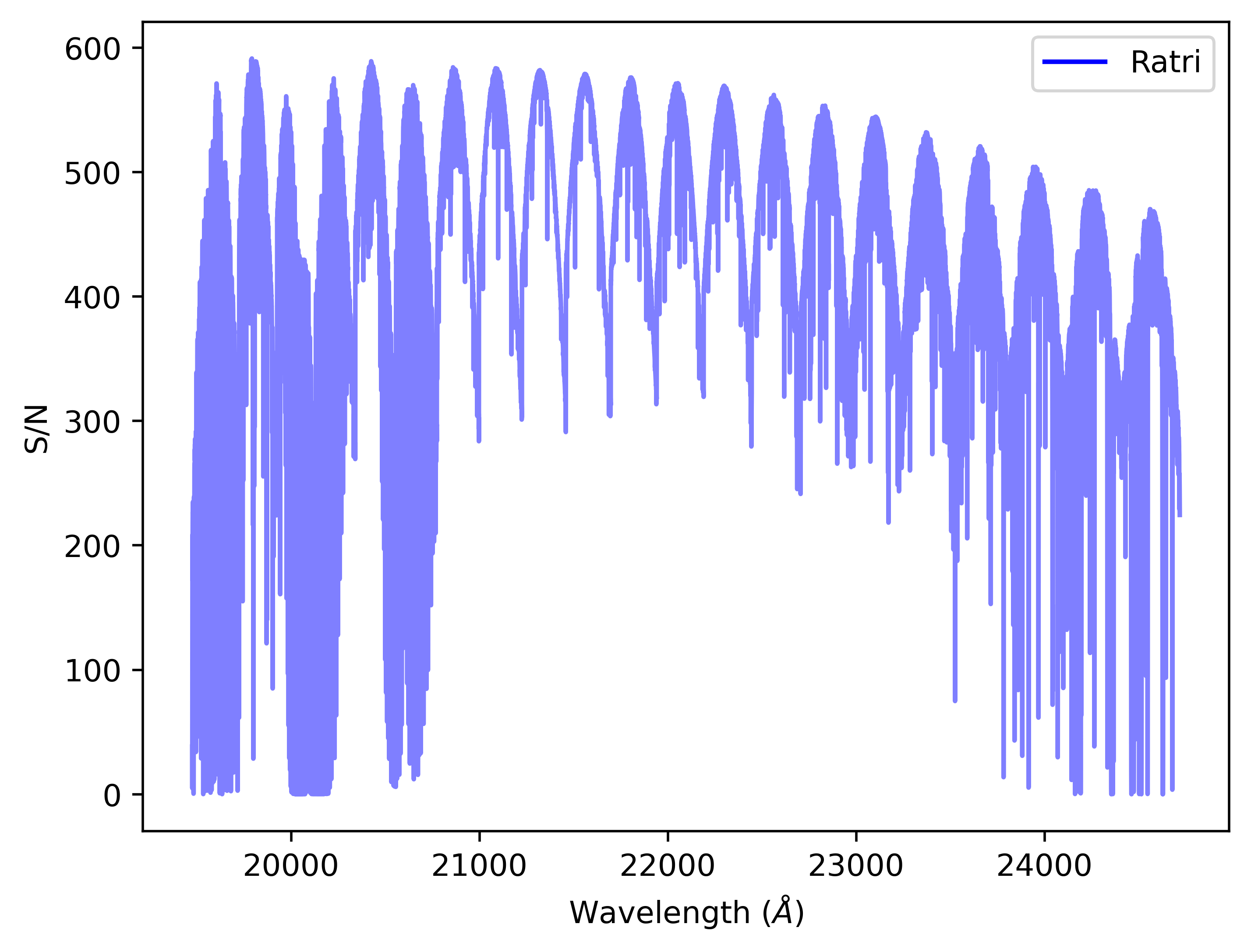}
  \caption{The SNR/resolution element for (top-left) Band UBV, (top right) Band RIZ, (bottom left) Band YJH, and (bottom right) Band K for V1298\,Tau obtained using \texttt{Ratri} \citep{dash2025detectability}. The values of the exposure time and airmass are the same as in Figure \ref{fig:ANDESexp}. While the SNR/resolution element values are slightly over-predicted (by about a factor of 1.1 at the peak values) compared to the values in Figure \ref{fig:ANDESexp}, the overall trend of the plots remains the same.}
  \label{fig:Ratriexp}
\end{figure*}

We compare the signal-to-noise ratio (SNR) per resolution element obtained from the ANDES exposure time calculator with the corresponding values produced by \texttt{Ratri} for the case of V1298\,Tau with exposure time of 60\,s and airmass of 1.2. Figures \ref{fig:ANDESexp} and \ref{fig:Ratriexp} show that, although \texttt{Ratri} slightly over-predicts the absolute SNR values, it reproduces the overall wavelength-dependent trend across all the available bands.

\clearpage

\section{An example 3-parameter retrieval for V1298\,Tau\,b} \label{andesratriret}

\begin{figure*}[h!]
\centering
  \includegraphics[width=0.85\columnwidth]{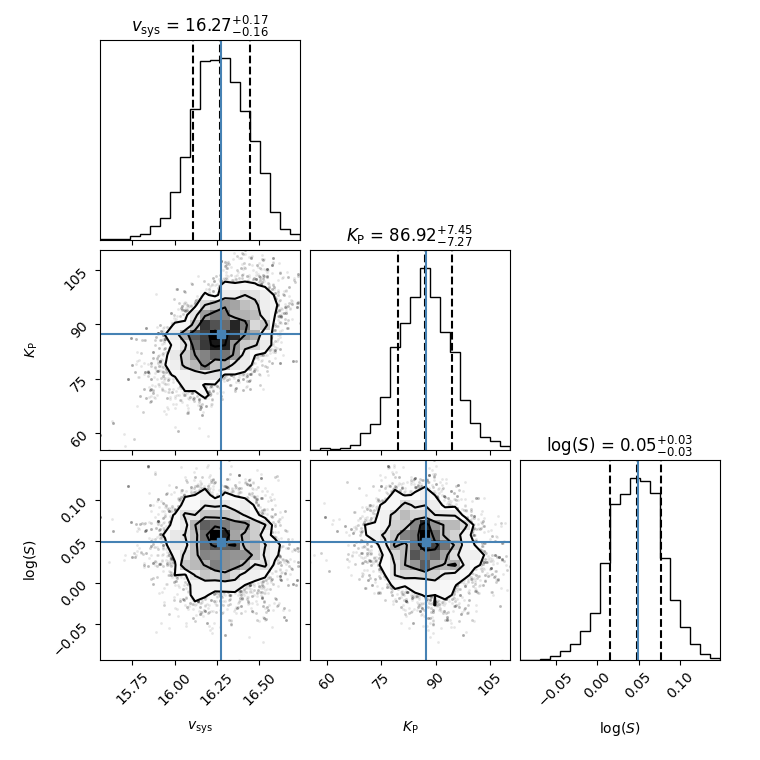}
  \caption{A three-component Markov Chain Monte Carlo (MCMC) retrieval of the orbital parameters and cross-correlated template model scale factor for an injected signal of V1298\,Tau\,b from 1000 iterations and using the two simulated nights (Nights 2 and 3) with out-of-transit phases from Section \ref{v1298taubobs}. The template models are reprocessed before cross-correlation. The injected values for $v_\mathrm{sys}$ and $K_\mathrm{P}$ fall within the 1$\sigma$ contours for their posteriors. The retrieved value of $\log(S)$ $\sim$ 0, which motivates our usage of $S=1$ in Equation \ref{logleq}. The python library \texttt{emcee} \citep{foreman2013emcee} was used to perform these MCMC retrievals, with uniform priors ($\mathcal{U}$(0.40), $\mathcal{U}$(0,200) and $\mathcal{U}$(-1,1) respectively) for all three parameters.}
  \label{fig:ret}
\end{figure*}

To illustrate the behaviour of the HRCCS likelihood framework used in this study, we show in Figure \ref{fig:ret} an example three-parameter retrieval for an injected V1298\,Tau\,b signal using the two simulated nights that include out-of-transit phases. The retrieval demonstrates that the injected values of ($v_{\mathrm{sys}}$) and ($K_\mathrm{p}$) are recovered within 1\,$\sigma$ of the posterior constraints, and that the preferred value of (log(S)) remains close to zero, which motivates our adoption of (S=1) in the main analysis.

\clearpage

\section{The effect of the application of \texttt{Upamana} on an injected signal template} \label{reprocesseffect}

\begin{figure*}[h!]
\centering
  \includegraphics[width=\columnwidth]{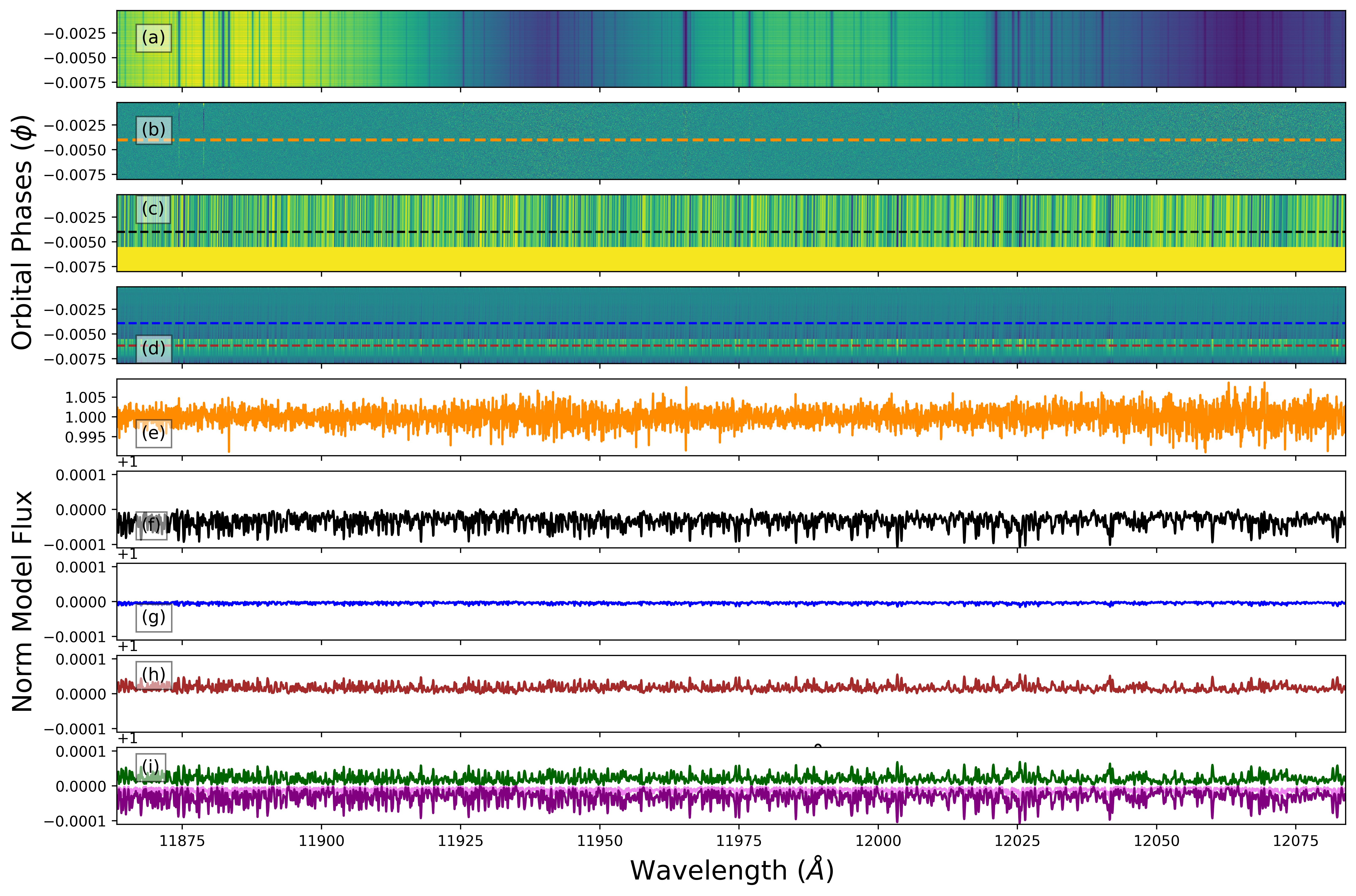}
  \caption{Visualisation of the effect of the SVD based detrending pipeline using in \texttt{Upamana} on an injected signal. Panel \textbf{(a)} shows one particular order for the dataset considered. Panel \textbf{(b)} shows the same order after detrending, with the cross-section in orange visualised in Panel \textbf{(e)}. Panel \textbf{(c)} shows the trace of the Doppler shifted signal injected in Panel (a), with the cross-section in black shown in panel \textbf{(f)}. The reprocessed model is shown in Panel \textbf{(d)} showcasing the effect of the detrending procedure on the injected signal. As evident, this is not similar to Panel (c), as can also be seen from the blue in-transit cross-section (in Panel \textbf{(g)}) being heavily attenuated compared to the signal expected. The cross-section in brown (Panel \textbf{(h)}) shows that there are artefacts in the out-of-transit phases, which are an attenuated mirror image of the actual signal. The maximum of the in-transit signal and out-of-transit signal are close to the ingress and are plotted in pink and green respectively in panel (i). Adding 1-maximum of out-of-transit to the maximum of in-transit signal recovers the actual model. Please see Appendix \ref{reprocesseffect} for more details.}
  \label{fig:reprocesseffectfig1}
\end{figure*}

\begin{figure*}[h!]
\centering
  \includegraphics[width=0.8\columnwidth]{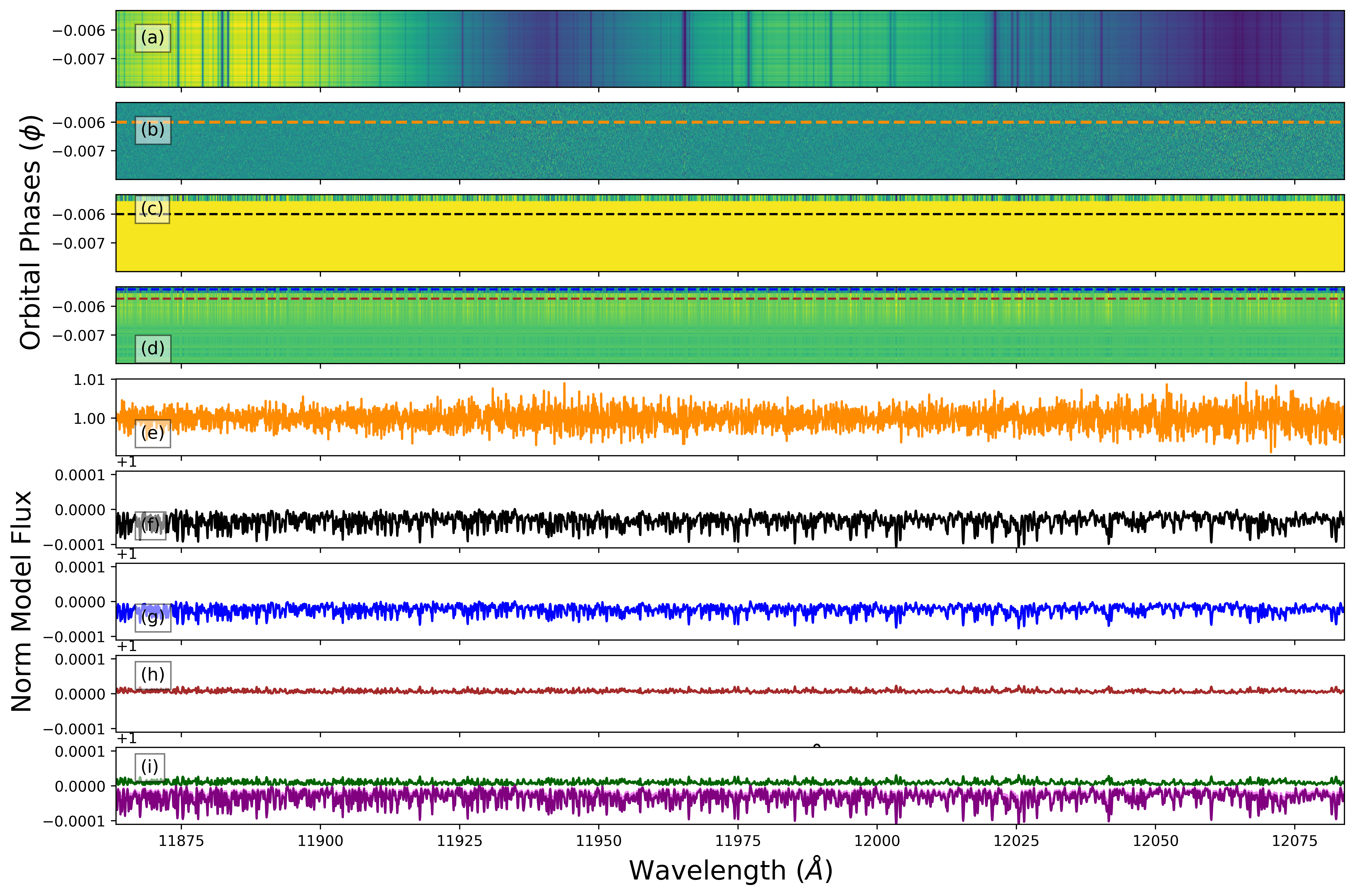}
  \includegraphics[width=0.8\columnwidth]{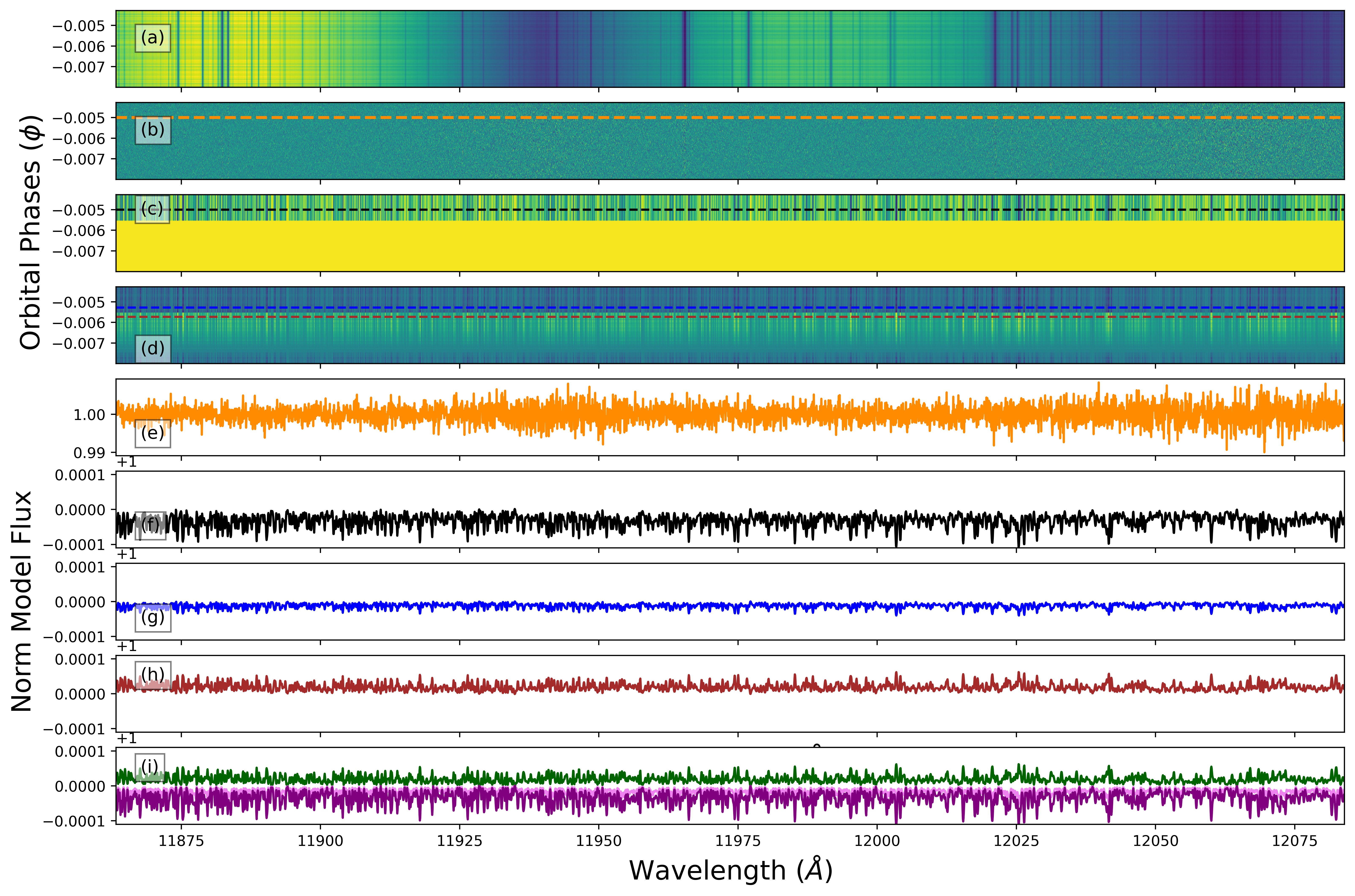}
  \caption{Same as in Figure \ref{fig:reprocesseffectfig1}, but now with the in-transit to out-of-transit spectra ratio being (top) 10\% and (bottom) 25\%. The combined in-transit and out-of-transit orbital trace is maximised for the 25\% case.}
  \label{fig:reprocesseffectfig2}
\end{figure*}

We illustrate the effect of application of the \texttt{Upamana} pipeline using spectral order 12 for the third simulated night for the case of V1298\,Tau\,b. Figure \ref{fig:reprocesseffectfig1} showcases a step-wise visualisation of the same, with Panel (a) being the original dataset, Panel (b) being the detrended dataset that is used for cross-correlation, with Panel (e) depicting the cross-section values at a particular orbital phase (in orange). As expected, they vary around 1.0, with the highest variation occurring in regions with the lowest flux values, and around the strongest telluric lines. Panel (c) represents the expected trace of the model injected into a noiseless dataset of ones, with Panel (f) depicting a cross-section at the same orbital phase as in Panel (b) (now in black). Panel (d) showcases the effect of injecting a model signal into the dataset of Panel (a) and then detrending by using $k=6$. It is already clear that the trace of the reprocessed model is very different from that in Panel (c), with the orbital trace rapidly disappearing further from the ingress phase, and the presence of out-of-transit artefacts before the ingress. To augment this, we depict the cross-sections at the in-transit orbital phase with the same value as in Panel (b) (in blue in Panel (g)) and an out-of-transit phase (in brown in Panel (h)). Panel (g) shows that the signal is almost completely attenuated at this orbital phase, but the out-of-transit artefact shows a distinct signature, which looks like an attenuated mirror image of the model in Panel (f). Now, we perform a test by taking the maximum value of the signal in transit (occurring just after ingress, cross-section in pink in Panel (i)) and adding it with 1-the maximum value of the out-of-transit artefacts (cross-section in green in Panel (i)). The resulting value in violet in Panel (i) is exactly equal to our injected model in Panel (f).
\\
\\
While the presence of the artefacts was already showcases in \citet{cheverall2024feasibility} and \citet{dash2024constraints}, we show here for the first time that these artefacts are actually an attenuated mirror image of the injected signal. The overall detection comes from an orbital trace composed of a severely attenuated in-orbital trace and an attenuated but still prominent out-of-orbit trace caused due to partitioning of an injected signal due to the SVD based detrending step. This is why we are able to get detections in the analysis of long-period exoplanets in this paper. A non-reprocessing scenario will not account for the out-of-transit artefacts and hence recover a much weakened detection. The constraints on $K_\mathrm{P}$ will also suffer as the orbital trace utilised to get the sum of CCFs is much shorter. Hence, reprocessing the models before cross-correlation and accounting for the out-of-transit artefacts is a necessity for long-period exoplanets like the candidates we use in this paper.
\\
\\
We also give a visual depiction of how these orbital traces change if the ratio of in-transit and out-of-transit (or ItoO in short) is changed in Figure \ref{fig:reprocesseffectfig2}. The ItoO ratio utilised in the top-left panel is 10\%, top-right panel is 25\%, bottom-left panel is 50\% and bottom-right panel is 80\%. Based on these, we recommend a value of ItoO between 25\% and 50\% (both inclusive) to maximise the planetary orbital trace from the in-transit and out-of-transit trace. This would help in pin-pointing the planetary signal better and in the event of a case with a much higher number of in-transit versus out-of-transit spectra, the out-of-transit traces might start getting reduced, thus weakening detection constraints (e.g. Night 2 in V1298\,Tau\,b).

\begin{figure*}[h!]
\centering
  \includegraphics[width=0.8\columnwidth]{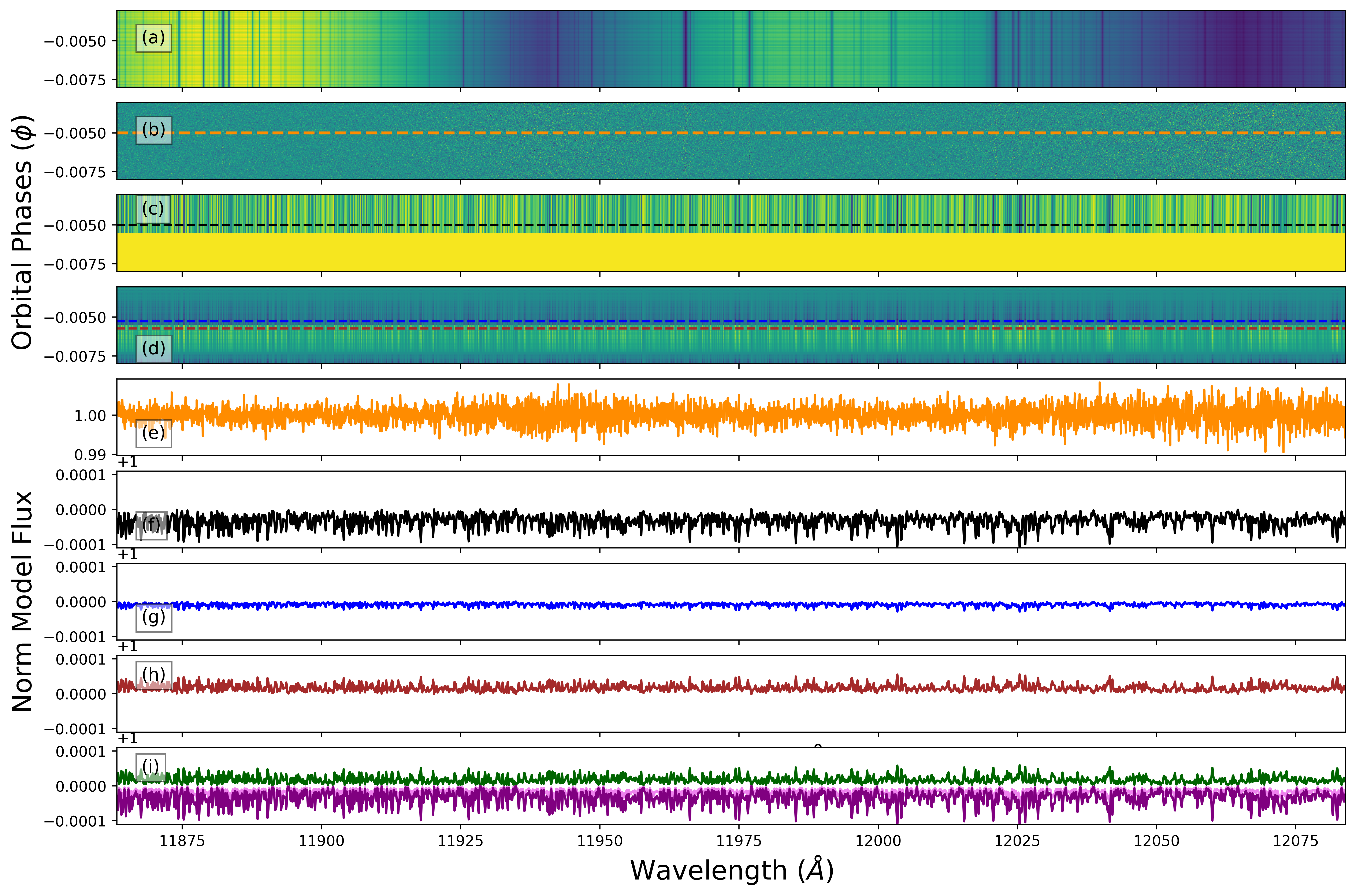}
  \includegraphics[width=0.8\columnwidth]{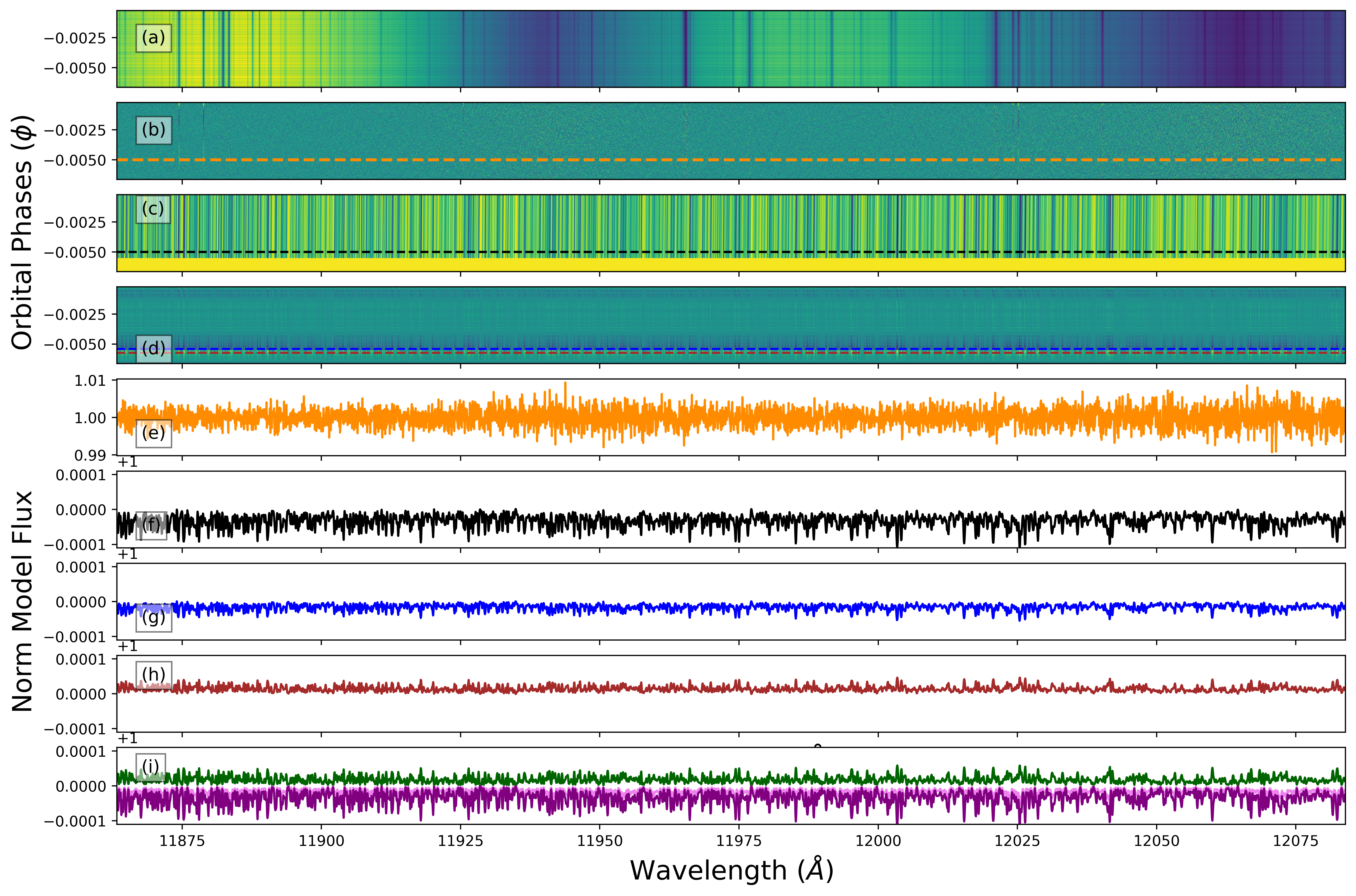}
  \caption{Same as in Figure \ref{fig:reprocesseffectfig2}, but now with the in-transit to out-of-transit spectra ratio being (top) 50\% and (bottom) 80\%. The combined in-transit and out-of-transit orbital trace is maximised for the 50\% case. From this figure and Figure \ref{fig:reprocesseffectfig2}, both 25\% to 50\% ratio seem the best candidates to maximise the orbital trail as both traces are heavily attenuated the further one moves from the point of ingress.}
  \label{fig:reprocesseffectfig3}
\end{figure*}

\clearpage
\bibliography{sample701}{}
\bibliographystyle{aasjournalv7}



\end{document}